\documentclass[preprint,pre,aps,showpacs,showkeys,amsmath]{revtex4}
\usepackage{graphicx}

\begin{document}

\title{Effect of Inter-Modular Connection on Fast Sparse Synchronization in Clustered Small-World Neural Networks}

\author{Sang-Yoon Kim}
\email{sykim@icn.re.kr}
\author{Woochang Lim}
\email{wclim@icn.re.kr}
\affiliation{Institute for Computational Neuroscience and Department of Science Education, Daegu National University of Education, Daegu 705-115, Korea}

\begin{abstract}
We consider a clustered network with small-world sub-networks of inhibitory fast spiking interneurons, and investigate the effect of inter-modular connection on emergence of fast sparsely synchronized rhythms by varying both the inter-modular coupling strength $J_{inter}$ and the average number of inter-modular links per interneuron $M_{syn}^{(inter)}$. In contrast to the case of non-clustered networks, two kinds of sparsely synchronized states such as modular and global synchronization are found. For the case of modular sparse synchronization, the population behavior reveals the modular structure, because the intra-modular dynamics of sub-networks make some mismatching. On the other hand, in the case of global sparse synchronization, the population behavior is globally identical, independently of the cluster structure, because the intra-modular dynamics of sub-networks make perfect matching. We introduce a realistic cross-correlation modularity measure, representing the matching-degree between the instantaneous sub-population spike rates of the sub-networks, and examine whether the sparse synchronization is global or modular. Depending on its magnitude, the inter-modular coupling strength $J_{inter}$ seems to play ``dual'' roles for the pacing between spikes in each sub-network. For large $J_{inter}$, due to strong inhibition it plays a destructive role to ``spoil'' the pacing between spikes, while for small $J_{inter}$ it plays a constructive role to ``favor'' the pacing between spikes. Through competition between the constructive and the destructive roles of $J_{inter}$, there exists an intermediate optimal $J_{inter}$ at which the pacing degree between spikes becomes maximal. In contrast, the average number of inter-modular links per interneuron $M_{syn}^{(inter)}$ seems to play a role just to favor the pacing between spikes. With increasing $M_{syn}^{(inter)}$, the pacing degree between spikes increases monotonically thanks to the increase in the degree of effectiveness of global communication between spikes. Furthermore, we employ the realistic sub- and whole-population order parameters, based on the instantaneous sub- and whole-population spike rates, to determine the threshold values for the synchronization-unsynchronization transition in the sub- and whole-populations, and the degrees of global and modular sparse synchronization are also measured in terms of the realistic sub- and whole-population statistical-mechanical spiking measures defined by considering both the occupation and the pacing degrees of spikes. It is expected that our results could have implications for the role of the brain plasticity in some functional behaviors associated with population synchronization.
\end{abstract}


\maketitle

\section{Introduction}
\label{sec:INT}
Recently, much attention has been paid to brain rhythms in health and disease \cite{Buz1,TW}. Particularly, we are interested in fast sparsely synchronized cortical rhythms which are associated with diverse cognitive functions such as sensory perception, feature integration, selective attention, and memory formation \cite{W_Review}. At the population level, local field potential recordings have been observed to show synchronous fast oscillations [e.g., gamma rhythm (30-100 Hz) and ultrafast sharp-wave ripple (100-200 Hz)], while individual neuronal recordings have been found to exhibit stochastic and intermittent spike discharges \cite{SS1,SS2,SS3,SS4,SS5,SS6,SS7}. Thus, single-cell firing activity differs markedly from the population oscillatory behavior. These sparsely synchronized rhythms are in contrast to fully synchronized rhythms. For the case of full synchronization, individual neurons fire regularly at the population frequency like the clock oscillators \cite{WB}. Hence, the fully synchronized oscillations may be well described by using the conventional coupled-oscillator model composed of suprathreshold spiking neurons above a threshold in the absence of noise or for weak noise \cite{Wang}. However, such coupled-oscillator models are not adequate for describing sparse synchronization because individual neurons fire stochastically at low rates like the Geiger counters. Brunel et al. in \cite{Sparse1,Sparse2,Sparse3,Sparse4,Sparse5,Sparse6} developed a framework appropriate for description of fast sparse synchronization by taking an opposite view from that of coupled oscillators. Under the condition of strong external noise, suprathreshold spiking neurons discharge irregular firings as Geiger counters, and then the population state becomes unsynchronized. However, when inhibitory recurrent feedback becomes sufficiently strong, this asynchronous state may be destabilized, and then a synchronous population state with irregular and intermittent individual discharges emerges. For this case, average total (external excitatory plus recurrent inhibitory) input current into individual neurons is subthreshold, but stochastic and intermittent firings are triggered when fluctuations (due to noise in external and recurrent inputs) cross a threshold. In this way, under the balance between strong external noise and strong recurrent inhibition, fast sparse synchronization was found to occur in networks of suprathreshold neurons \cite{Sparse1,Sparse2,Sparse3,Sparse4,Sparse5,Sparse6}. Similar sparsely synchronized rhythms were also found to appear via cooperation of noise-induced spikings of subthreshold neurons (which can not fire spontaneously without noise) \cite{Kim1,Kim2,Kim3}. However, in contrast to the above works on suprathreshold neurons, sparse synchronization for the case of subthreshold neurons has been found to appear under relatively weak external noise and recurrent inhibition, and the sparsely synchronized rhythms were also found to be slow when compared with the suprathreshold case.

In this paper, we are concerned about emergence of fast sparsely synchronized rhythms in an ensemble of suprathreshold neurons, as in the previous works of Brunel et al. \cite{Sparse1,Sparse2,Sparse3,Sparse4,Sparse5,Sparse6} where both random and global synaptic couplings were considered. However, connection architecture of the real brain has been found to have complex topology which is neither regular nor random \cite{Sporns,Buz2,CN1,CN2,CN3,CN4,CN5,CN6,CN7}. Particularly, mammalian (e.g., cat and macaque) brain anatomical networks and human brain functional (fMRI) networks have been revealed to have a modular structure composed of relatively sparsely linked clusters with spatial localization, as in social, technological, and biological complex networks \cite{MN,MN1,MN2,MN3,MN4,FMN1,FMN2,FMN3}. Within each cluster, nodes are highly inter-connected and exhibit similar connectional and functional features. This clustered organization of the brain network reveals the anatomical substrate for segregation which refers to the subdivision of the brain into regions specialized in particular functional tasks \cite{MN5,MN6,MN7}. This segregation allows the brain to process information in parallel, simultaneously by distinct populations of neurons. However, for emergence of a coherent perception and comprehensive understanding of the environment as a whole, specialized information of different modalities and features should be integrated. This integration refers to the capacity of a system to collect information of different nature and combine it to produce new useful information. For example, sensory perception requires the binding of the features of a receptive field (e.g., color, orientation, and position of a visual object). In this way, brain connectivity should be organized into a balance between segregation (specialization) and integration (binding) \cite{MN8,MN9,MN10,MN11,MN12}. Here, in our brain network we take into consideration the modular structure of the real brain. For modeling the modular structure of real brain, we consider a clustered network composed of sparsely connected sub-networks. The sub-networks stand for the modules (clusters) of the brain. As is also known, the connection structure in each module of the real brain reveals complex topology such as small-worldness and scale-freeness \cite{Sporns,Buz2,CN1,CN2,CN3,CN4,CN5,CN6,CN7}. Here, each sub-network (representing a cluster) is modeled as the Watts-Strogatz  small-world network which interpolates between the regular lattice with high clustering (i.e., high cliquishness of a typical neighborhood) and the random graph with short path length (i.e., average short separation between two neurons represented by average number of synapses between two neurons along the minimal path) by varying the rewiring probability $p_{rewiring}$ from local to long-range connections; $p_{rewiring}=0$ and 1 correspond to the regular lattice and the random graph, respectively \cite{SWN1,SWN2,SWN3}. The Watts-Strogatz model for the small-world sub-network may be regarded as a cluster-friendly extension of the random network by reconciling the six degrees of separation (small-worldness) \cite{SDS1,SDS2} with the circle of friends (clustering). Many recent works on various subjects of neurodynamics have been done in small-world networks with predominantly local connections and rare long-distance connections \cite{SW2,SW3,SW4,SW5,SW6,SW7,SW8,SW9,SW10,SW11,SW12,SW13}. Effect of this small-world connectivity on fast sparse synchronization has also been studied in our recent work \cite{Kim}.

We note that real brain networks, consisting of sparsely inter-connected modules, are far more complex than minimal non-modular models such as small-world and scale-free networks. The main purpose of our study is to investigate emergence of sparsely synchronized rhythms in more realistic modular networks. Clustered neural networks composed of regular, small-world, and scale-free sub-networks have been employed for study on several subjects of neurodynamics \cite{MN5,MN6,MN7,MN13,MN14,MN15,MN16,MN17}. For our aim, we consider a clustered network with small-world sub-networks of inhibitory spiking neurons, and investigate the effect of inter-modular connection on emergence of fast sparsely synchronized rhythms. In the absence of inter-modular coupling, we consider three cases for the intra-modular dynamics in sub-networks: (1) synchronized in all identical sub-networks, (2) unsynchronized in all identical sub-networks, and (3) synchronized/unsynchronized in non-identical sub-networks. For each case, we study the population states by changing both the inter-modular coupling strength $J_{inter}$ and the average number of inter-modular links per interneuron $M_{syn}^{(inter)}$. Consequently, two kinds of sparse synchronization such as modular and global synchronization are found, in contrast to the case of non-modular networks \cite{Sparse1,Sparse2,Sparse3,Sparse4,Sparse5,Sparse6,Kim,Kim-SFN}. For the case of modular sparse synchronization, the population behavior reveals the modular structure, and hence the degree of sparse synchronization in the whole population becomes less than that in the sub-networks, because the intra-modular dynamics of sub-networks make some mismatching. In contrast, for the case of global sparse synchronization, the population behavior is globally identical, independently of the cluster structure, because the intra-modular dynamics of sub-networks make perfect matching. These modular and global synchronization may be well visualized in the raster plots of spikes. For the case of synchronization, synchronous ``stripes'' (composed of spikes and indicating population synchronization) appear successively in the raster plots, while spikes are completely scattered (without forming any stripes) in the case of unsynchronization. Synchronization pacing (representing the smearing of spiking stripes: less smearing, better pacing) varies depending on $J_{inter}$. For large $J_{inter}$ it plays a destructive role to spoil the pacing between sparse spikes, because of strong inhibition. Hence, when passing a large threshold a transition to unsynchronization occurs. However, for small $J_{inter}$ it plays a constructive role to favor the pacing between spikes in each sub-network. Hence, via competition between the constructive and the destructive roles of $J_{inter}$, there appears an intermediate optimal $J_{inter}$ at which the pacing degree between spikes becomes maximal. In this way, $J_{inter}$  plays dual roles for the pacing between spikes in each sub-network, depending on its magnitude. On the other hand, the average number of inter-modular links per interneuron $M_{syn}^{(inter)}$ plays a role just to favor the pacing between spikes. As $M_{syn}^{(inter)}$ is increased, the pacing degree between spikes increases monotonically due to the increase in the degree of effectiveness of global communication between spikes. To make characterization of the synchronization-unsynchronization transitions in the sub- and whole-populations, we employ the realistic sub- and whole-population order parameters, based on the instantaneous sub- and whole-population spike rates \cite{RM}. Moreover, we introduce a realistic cross-correlation modularity measure, representing the matching-degree between the instantaneous sub-population spike rates of  sub-networks, and examine whether the sparse synchronization is global or modular. The degrees of modular and global sparse synchronization are also measured in terms of the realistic sub- and whole-population statistical-mechanical spiking measures defined by considering both the occupation and the pacing degrees of the spikes \cite{RM}.

This paper is organized as follows. In Sec.~\ref{sec:CSWN}, we describe a clustered network with small-world sub-networks of inhibitory fast spiking (FS) interneurons, and then the governing equations for the population dynamics
are given. Detailed explanations on methods for characterization of individual and population states in clustered networks are also given in Sec.~\ref{sec:Method}. Then, in Sec.~\ref{sec:EIC} we investigate the effect of inter-modular connection on emergence of fast sparsely synchronized rhythms by varying both $J_{inter}$ and $M_{syn}^{(inter)}$. Finally, a summary is given in Section \ref{sec:SUM}.

\section{Clustered Small-World Network of Inhibitory FS Izhikevich Interneurons}
\label{sec:CSWN}
In this section, we first describe our cluster network composed of $M (=3)$ small-world sub-networks, each of which contains $L$ FS Izhikevich interneurons in the subsection \ref{subsec:CSWN}. Then, the governing equations for the population dynamics in the clustered small-world network are given in the subsection \ref{subsec:GE}.

\subsection{Clustered Small-World Networks}
\label{subsec:CSWN}
\begin{figure}
\includegraphics[width=0.8\columnwidth]{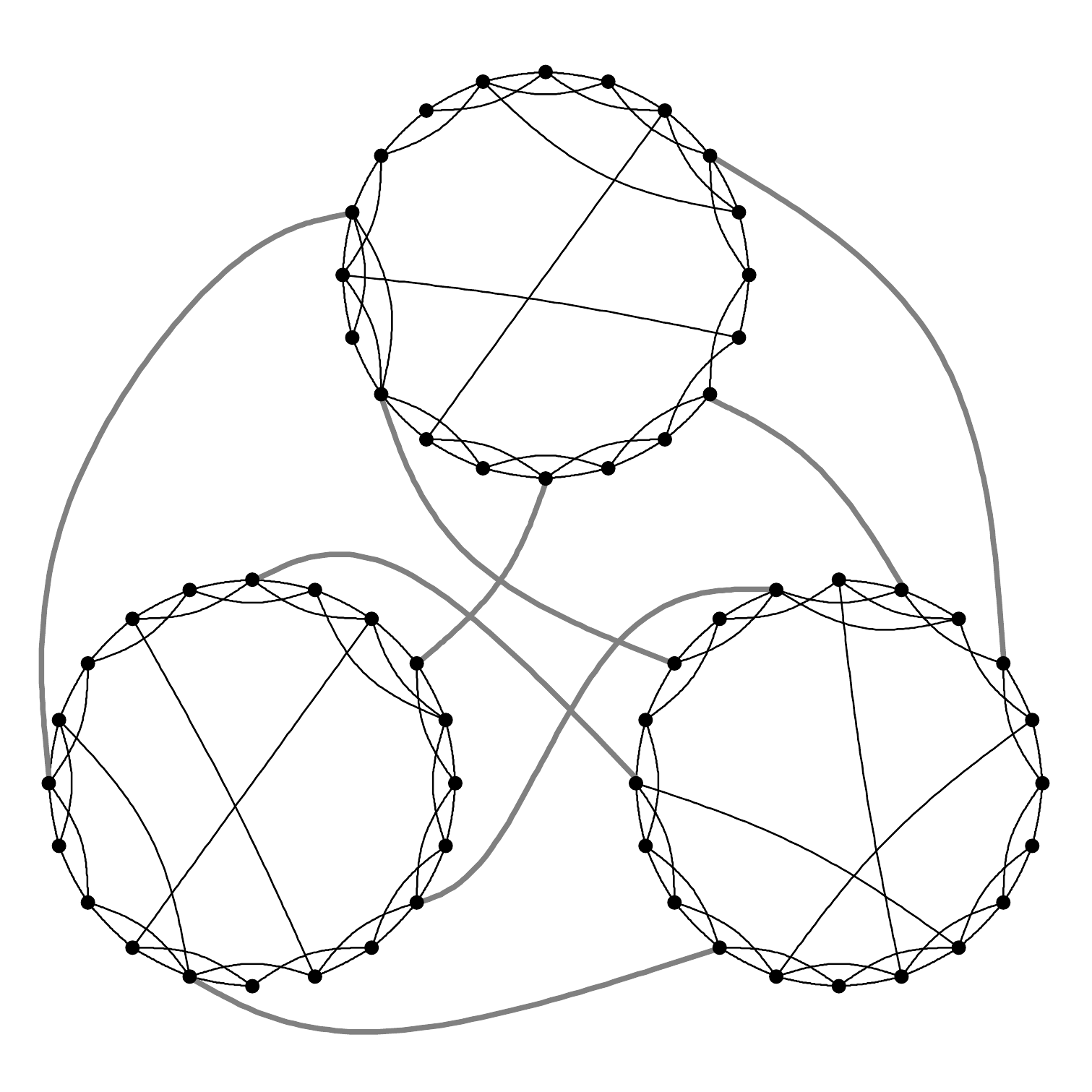}
\caption{Schematic representation of the clustered small-world network topology. The whole network consists of $M(=3)$ Watts-Strogatz small-world
sub-networks, each of them containing $L(=20)$ interneurons. Within each sub-network, the average number of intra-modular synaptic inputs per interneuron is
$M_{syn}^{(intra)} (=4)$, while there are 8 sparse random inter-modular connections (denoted by heavy lines) between the sub-networks.
}
\label{fig:CSWN}
\end{figure}

We consider a clustered network with $M$ $(=3)$ small-world sub-networks. Each small-world sub-network consists of $L$ inhibitory interneurons equidistantly placed on a one-dimensional ring of radius $L/ 2 \pi$. For illustrative purpose, an example of the clustered network topology is shown in Fig.~\ref{fig:CSWN}. Each of the three sub-networks, consisting of $L=20$ interneurons, is modeled as the Watts-Strogatz small-world network which interpolates between the regular lattice and the random graph by varying the rewiring probability $p_{rewiring}$ from local to long-range connections \cite{SWN1,SWN2,SWN3}. We start from the case of $p_{rewiring}=0$, corresponding to a directed regular ring lattice where each interneuron is coupled to its first $M_{syn}^{(intra)}$ $(=4)$ neighbors ($M_{syn}^{(intra)}/2$ on either side) via outward synapses. Then, we rewire each outward connection at random with probability $p_{rewiring}$ such that self-connections and duplicate connections are excluded, and the value of $p_{rewiring}$ is 0.25 for the case of Fig.~\ref{fig:CSWN}. Within each small-world sub-network, the average number of intra-modular synaptic inputs per interneuron is $M_{syn}^{(intra)}$, while there exist 8 sparse random inter-modular links between small-world sub-networks.

\subsection{Governing Equations for The Population Dynamics}
\label{subsec:GE}
As an element in our clustered small-world network, we choose the FS Izhikevich interneuron model which is not only biologically plausible, but also computationally efficient \cite{Izhi1,Izhi2,Izhi3,Izhi4}.
We consider the clustered network composed of $M (=3)$ small-world subnetworks, each of which consists of $L$ FS interneurons; $L=10^3$, except for the case of order parameters and spatial cross-correlation functions.
The following equations (\ref{eq:CIZA})-(\ref{eq:CIZF}) govern the population dynamics
in the clustered small-world network:
\begin{eqnarray}
C\frac{dv_{i}^{(I)}}{dt} &=& k (v_{i}^{(I)} - v_r) (v_{i}^{(I)} - v_t) - u_{i}^{(I)} +I_{DC} +D \xi_{i}^{(I)} -I_{I,i}^{(intra,syn)} -I_{I,i}^{(inter,syn)}, \label{eq:CIZA} \\
\frac{du_{i}^{(I)}}{dt} &=& a \{ U(v_{i}^{(I)}) - u_{i}^{(I)} \},  \;\;\; i=1, \cdots, L, \;\;\; I=1, \cdots, M, \label{eq:CIZB}
\end{eqnarray}
with the auxiliary after-spike resetting:
\begin{equation}
{\rm if~} v_{i}^{(I)} \geq v_p,~ {\rm then~} v_{i}^{(I)} \leftarrow c~ {\rm and~} u_{i}^{(I)} \leftarrow u^{(I)}_{i} + d, \label{eq:RS}
\end{equation}
where
\begin{eqnarray}
U(v) &=& \left\{ \begin{array}{l} 0 {\rm ~for~} v<v_b \\ b(v - v_b)^3 {\rm ~for~} v \ge v_b \end{array} \right. , \label{eq:CIZC} \\
I_{I,i}^{(intra,syn)} &=& \frac{J_{intra}}{d_{I,i}^{intra}} \sum_{j=1(\ne i)}^L w_{ij}^{(I,I)} s_j^{(I)}(t) (v_i^{(I)} - V_{syn}), \label{eq:CIZD}\\
I_{I,i}^{(inter,syn)} &=& \frac{J_{inter}}{d_{I,i}^{inter}} \sum_{J=1(\ne I)}^M \sum_{j=1}^L w_{ij}^{(I,J)} s_j^{(J)}(t) (v_i^{(I)} - V_{syn}), \label{eq:CIZE} \\
s_j^{(I)} (t) &=& \sum_{f=1}^{F^{(I)}_j} E(t-t_f^{(I,j)}-\tau_l);~E(t) = \frac{1}{\tau_d - \tau_r} (e^{-t/\tau_d} - e^{-t/\tau_r}) \Theta(t). \label{eq:CIZF}
\end{eqnarray}
Here, $v_i^{(I)}(t)$ and $u_i^{(I)}(t)$ are the state variables of the $i$th interneuron in the $I$th sub-network at a time $t$ which represent the membrane potential and the recovery current, respectively.
These membrane potential and the recovery variable, $v_i^{(I)}(t)$ and $u_i^{(I)}(t)$, are reset according to Eq.~(\ref{eq:RS}) when $v_i^{(I)}(t)$ reaches its cutoff value $v_p$.
$C$, $v_r$, and $v_t$ in Eq.~(\ref{eq:CIZA}) are the membrane capacitance, the resting membrane potential, and the instantaneous threshold potential, respectively.
The parameter values used in our computations are listed in Table \ref{tab:Parm}. More details on the FS Izhikevich interneuron model, the external stimulus to the FS interneuron, the intra-modular and the inter-modular synaptic currents, and numerical integration of the governing equations are given in the following subsubsections.

\subsubsection{FS Izhikevich Interneuron Model}
\label{subsubsec:Izhi}
The Izhikevich model matches neuronal dynamics by tuning the parameters $(k, a, b, c, d)$ instead of matching neuronal electrophysiology, in contrast to Hodgkin-Huxley-type conductance-based models.
The parameters $k$ and $b$ are related to the neuron's rheobase and input resistance, and $a$, $c$, and $d$ are the recovery time constant, the after-spike reset value of $v$, and the total amount of outward minus inward currents during the spike and affecting the after-spike behavior (i.e., after-spike jump value of $u$), respectively. Depending on the values of these parameters, the Izhikevich neuron model may exhibit 20 of the most prominent neuro-computational features of cortical neurons \cite{Izhi1,Izhi2,Izhi3,Izhi4}. Here, we use the parameter values for the FS interneurons in the layer 5 rat visual cortex, which are listed in the 1st item of Table \ref{tab:Parm}.

\subsubsection{External Stimulus to The FS Izhikevich Interneuron}
\label{subsubsec:Sti}
Each Izhikevich interneuron is stimulated by both a common DC current $I_{DC}$ and an independent Gaussian white noise $\xi_i^{(I)}$, as shown in the 3rd and the 4th terms in Eq.~(\ref{eq:CIZA}). The Gaussian white noise satisfies $\langle \xi_i^{(I)}(t) \rangle =0$ and $\langle \xi_i^{(I)}(t)~\xi_j^{(J)}(t') \rangle = \delta_{IJ}~\delta_{ij}~\delta(t-t')$, where $\langle\cdots\rangle$ denotes the ensemble average. Here the Gaussian noise $\xi$ is a parametric one that randomly perturbs the strength of the applied current $I_{DC}$, and its intensity is controlled by the parameter $D$. For $D=0$, the Izhikevich interneuron exhibits a jump from a resting state to a spiking state via subcritical Hopf bifurcation for a higher threshold $I_{DC,h} (\simeq 73.7)$ by absorbing an unstable limit cycle born via a fold limit cycle bifurcation at a lower threshold $I_{DC,l} (\simeq 72.8)$. Therefore, the Izhikevich interneuron shows type-II excitability since it begins to fire with a non-zero frequency \cite{Ex1,Ex2}. With increasing $I_{DC}$ from $I_{DC,h}$, the mean firing rate $f$ increases monotonically.
The values of $I_{DC}$ and $D$ used in this paper are given in the 2nd item of Table \ref{tab:Parm}.

\subsubsection{Intra-modular and Inter-modular Synaptic Currents}
\label{subsubsec:Syn}
The last two terms in Eq.~(\ref{eq:CIZA}) represent the intra- and the inter-modular synaptic couplings of inhibitory FS interneurons. $I_{I,i}^{(intra,syn)}$ and $I_{I,i}^{(inter,syn)}$ of Eqs.~(\ref{eq:CIZD}) and (\ref{eq:CIZE}) represent the intra- and the inter-modular synaptic currents injected into the $i$th neuron in the $I$th sub-network, respectively. The synaptic connectivity is given by the connection weight matrix $W$ (=$\{ w_{ij}^{(I,J)} \}$) where  $w_{ij}^{(I,J)}=1$ if the neuron $j$ in the $J$th sub-network is presynaptic to the neuron $i$ in the $I$th sub-network; otherwise, $w_{ij}^{(I,J)}=0$. Here, the intra-modular synaptic connection is modeled in terms of the Watts-Strogatz small-world network. Then, the in-degree of the $i$th neuron in the $I$th sub-network for the intra-modular synaptic connection, $d_{I,i}^{intra}$ (i.e., the number of intra-modular synaptic inputs to the neuron $i$ in the $I$th sub-network) is given by $d_{I,i}^{intra} = \sum_{j=1(\ne i)}^L w_{ij}^{(I,I)}$. For this intra-modular case, the average number of intra-modular synaptic inputs per neuron is $M_{syn}^{(intra)} = \frac{1}{M \cdot L} \sum_{I=1}^{M} \sum_{i=1}^{L} d_{I,i}^{intra}$. In contrast to the intra-modular connection, the inter-modular synaptic connection is given randomly. Then, the in-degree of the $i$th neuron in the $I$th sub-network for the inter-modular synaptic connection, $d_{I,i}^{inter}$ (i.e., the number of inter-modular synaptic inputs to the neuron $i$ in the $I$th sub-network) is given by $d_{I,i}^{inter} =  \sum_{J=1(\ne I)}^M \sum_{j=1}^L w_{ij}^{(I,J)}$. In the inter-modular case, the average number of inter-modular synaptic inputs per neuron is $M_{syn}^{(inter)} = \frac{1}{M \cdot L} \sum_{I=1}^{M} \sum_{i=1}^{L} d_{I,i}^{inter}$, and these inter-modular links are randomly connected with the inter-modular connection probability $p_{inter} = \frac{M_{syn}^{(inter)}}{(M-1) \cdot L}$. Compared to the intra-modular connections, the inter-modular connections are sparse (i.e.,
$M_{syn}^{(inter)} < M_{syn}^{(intra)}$). The fraction of open synaptic ion channels at time $t$ is denoted by $s(t)$. The time course of $s_j^{(I)}(t)$ of the $j$th neuron in the $I$th sub-network is given by a sum of delayed double-exponential functions $E(t-t_f^{(I,j)}-\tau_l)$ [see Eq.~(\ref{eq:CIZF})], where $\tau_l$ is the synaptic delay, and $t_f^{(I,j)}$ and $F_j^{(I)}$ are the $f$th spiking time and the total number of spikes of the $j$th neuron in the $I$th sub-network at time $t$, respectively. Here, $E(t)$ [which corresponds to contribution of a presynaptic spike occurring at time $0$ to $s(t)$ in the absence of synaptic delay] is controlled by the two synaptic time constants: synaptic rise time $\tau_r$ and decay time $\tau_d$, and $\Theta(t)$ is the Heaviside step function: $\Theta(t)=1$ for $t \geq 0$ and 0 for $t <0$. The intra- and the inter-modular synaptic coupling strengths are controlled by the parameters $J_{intra}$ and $J_{inter}$, respectively and $V_{syn}$ is the synaptic reversal potential. For the inhibitory GABAergic synapse (involving the $\rm{GABA_A}$ receptors), the values of $\tau_l$, $\tau_r$, $\tau_d$, and $V_{syn}$ are listed in the 3rd item of Table \ref{tab:Parm}.

\subsubsection{Numerical Method}
\label{subsubsec:NM}
Numerical integration of stochastic differential Eqs.~(\ref{eq:CIZA})-(\ref{eq:CIZB}) is done by employing the Heun method \cite{SDE} with the time step $\Delta t=0.01$ ms. For each realization of the stochastic process, we choose a random initial point $[v_i^{(I)}(0),u_i^{(I)}(0)]$ for the $i$th $(i=1,\dots, N)$ neuron in the $I$th sub-network with uniform probability in the range of $v_i^{(I)}(0) \in (-50,-45)$ and $u_i^{(I)}(0) \in (10,15)$.

\section{Methods for Characterization of Individual And Population States in Clustered Networks}
\label{sec:Method}
In the following subsections, we explain methods used to characterize individual and population states in clustered networks. Particularly, emergence of population synchronization and its degree are characterized by employing realistic measures, based on instantaneous sub- and whole-population spike rates \cite{RM}. Furthermore, we introduce a realistic cross-correlation modularity measure, denoting the matching-degree between the instantaneous sub-population spike rates to examine whether the population synchronization is global or modular.

\subsection{Characterization of Individual Firing Behaviors}
\label{subsec:IB}
Firing behaviors of individual interneurons are characterized in terms of the inter-spike interval (ISI) histogram and the mean firing rate (MFR) distribution. The ISI histogram is composed of $5 \times 10^4$ ISIs (obtained from all the interneurons), and the bin size for the histogram is 0.5 ms. The MFR for each interneuron is calculated by following the membrane potential during the averaging time of $2 \times 10^4$ ms after discarding the transient time of $10^3$ ms, and the bin size for the histogram is 2 Hz.

\subsection{Sub- and Whole-population Variables}
\label{subsec:SWV}
In computational neuroscience, an ensemble-averaged sub-population potential $V_s^{(I)} (t)$ for the $I$th sub-network ($I=1,2,3$), containing $L$ FS Izhikevich interneurons,
\begin{equation}
V_s^{(I)} (t) = \frac {1} {L} \sum_{i=1}^{L} v_i^{(I)}(t),
\label{eq:SPP}
\end{equation}
and an ensemble-averaged whole-population potential for the whole network with $M$ (=3) sub-networks,
\begin{equation}
V_w (t) = \frac {1} {M} \sum_{I=1}^{M} V_s^{(I)}(t),
\label{eq:WPP}
\end{equation}
are often used for describing emergence of population neural synchronization in the sub- and the whole-populations, respectively (e.g., sparse synchronization in a population of subthreshold neurons was described in terms of an ensemble-averaged global potential \cite{Kim1,Kim2,Kim3}). However, to directly obtain $V_s^{(I)}(t)$ and $V_w(t)$ in real experiments is very difficult. To overcome this difficulty, instead of $V_s^{(I)}(t)$ and $V_w(t)$, we use an experimentally-obtainable instantaneous sub- and whole-population spike rates which are often used as collective quantities showing sub- and whole-population behaviors \cite{W_Review,Sparse1,Sparse2,Sparse3,Sparse4,Sparse5,Sparse6,Kim,Kim-SFN,RM}. The instantaneous sub-population spike rate (ISPSR) $R_s^{(I)}(t)$ is obtained from the raster plot of neural spikes which is a collection of spike trains of individual neurons in the $I$th sub-population. Such raster plots of spikes, where sub-population spike synchronization may be well visualized, are fundamental data in experimental neuroscience. For the synchronous case, ``stripes" (composed of spikes and indicating sub-population synchronization) are found to be formed in the raster plot. Hence, for a synchronous case, an oscillating ISPSR appears, while for an unsynchronized case the ISPSR is nearly stationary. To obtain a smooth ISPSR, we employ the kernel density estimation (kernel smoother) \cite{Kernel}. Each spike in the raster plot is convoluted (or blurred) with a kernel function $K_h(t)$ to obtain a smooth estimate of ISPSR for the $I$th sub-network, $R_s^{(I)}(t)$:
\begin{equation}
R_s^{(I)}(t) = \frac{1}{L} \sum_{i=1}^{L} \sum_{s=1}^{n_i^{(I)}} K_h (t-t_{s}^{(I,i)}),
\label{eq:ISPSR}
\end{equation}
where $t_{s}^{(I,i)}$ is the $s$th spiking time of the $i$th neuron in the $I$th sub-network, $n_i^{(I)}$ is the total number of spikes for the $i$th neuron in the $I$th sub-network, and we use a Gaussian kernel function of band width $h$:
\begin{equation}
K_h (t) = \frac{1}{\sqrt{2\pi}h} e^{-t^2 / 2h^2}, ~~~~ -\infty < t < \infty.
\label{eq:Gaussian}
\end{equation}
Throughout the paper, the band width of the Gaussian kernel estimate is $h=1$ ms. Then, the instantaneous whole-population spike rate (IWPSR) kernel estimate $R_w(t)$ for the whole population is given by an average of the ISPSR kernel estimates of the $M(=3)$ sub-populations:
\begin{equation}
R_w (t) = \frac {1} {M} \sum_{I=1}^{M} R_s^{(I)}(t).
\label{eq:IWPSR}
\end{equation}
Moreover, for the synchronous case, the sub-population frequency $f_p^{(I)}$ of the regularly-oscillating ISPSR $R_s^{(I)}(t)$ may be obtained from the one-sided power spectrum of $\Delta R_s^{(I)}(t)$ $[= R_s^{(I)}(t) - \overline{R_s^{(I)}(t)}]$ with the mean-squared amplitude normalization. The number of data for the power spectrum is $2^{13}$, and the overline denotes the time average.

\subsection{Sub- and Whole-population Order Parameters}
\label{subsec:SWO}
As is well known, a conventional order parameter, based on the ensemble-averaged global potential, is often used for describing transition from synchronization to unsynchronization in computational neuroscience \cite{Kim1,Kim2,Kim3,Order1,Order2,Order3}. Recently, instead of the global potential, we used an experimentally-obtainable instantaneous population spike rate kernel estimate, and developed a realistic order parameter for the case of the non-modular networks, which may be applicable in both the computational and the experimental neuroscience \cite{RM}. For the case of modular networks, the mean square deviation of the ISPSR kernel estimate $R_s^{(I)}(t)$ for the $I$th sub-network ($I$=1, 2, 3),
\begin{equation}
{\cal{O}}_s^{(I)} \equiv \overline{(R_s^{(I)}(t) - \overline{R_s^{(I)}(t)})^2},
\label{eq:SOrder}
\end{equation}
and the mean square deviation of the IWPSR kernel estimate $R_w(t)$ for the whole network,
\begin{equation}
{\cal{O}}_w \equiv \overline{(R_w(t) - \overline{R_w(t)})^2},
\label{eq:WOrder}
\end{equation}
play the role of realistic sub- and whole-population order parameters ${\cal{O}}_s^{(I)}$ and ${\cal{O}}_w$ to determine a threshold for the synchronization-unsynchronization transition, where the overbar represents the time average. Here, each order parameter is obtained through average over 20 realizations, and the averaging time for the calculation of the order parameter in each realization is $4 \times 10^3$ ms.
Then, the order parameters ${\cal{O}}_s^{(I)}$ and ${\cal{O}}_w$, representing the time-averaged fluctuations of $R_s^{(I)}(t)$ and $R_w(t)$, approach non-zero (zero) limit values for the synchronized (unsynchronized) state in the thermodynamic limit of $L \rightarrow \infty$. These order parameters may be regarded as thermodynamic measures because they concern just the macroscopic ISPSR and IWPSR kernel estimates $R_s^{(I)}(t)$ and $R_w(t)$ without any consideration between the macroscopic ISPSR and IWPSR kernel estimates and microscopic individual spikes.

\subsection{Spatial Cross-correlation Functions}
\label{subsec:SCF}
To further understand the synchronization-unsynchronization transition, we consider the ``microscopic'' dynamical cross-correlations between neuronal pairs. For obtaining dynamical pair cross-correlations, each spike train of the $i$th neuron in the $I$th sub-network is convoluted with a Gaussian kernel function $K_h(t)$ of band width $h$ to get a smooth estimate of instantaneous individual spike rate (IISR) $r_i^{(I)}(t)$:
\begin{equation}
r_i^{(I)}(t) = \sum_{s=1}^{n_i^{(I)}} K_h (t-t_{s}^{(I,i)}),
\label{eq:IISR}
\end{equation}
where $t_{s}^{(I,i)}$ is the $s$th spiking time of the $i$th neuron in the $I$th sub-network, $n_i^{(I)}$ is the total number of spikes for the $i$th neuron, and $K_h(t)$ is given in Eq.~(\ref{eq:Gaussian}). Then, the normalized temporal cross-correlation function $C_{i,j}^{(I)}(\tau)$ between the IISR kernel estimates $r_i^{(I)}(t)$ and $r_j^{(I)}(t)$ of the $(i,j)$ neuronal pair
in the $I$th sub-network is given by:
\begin{equation}
C_{i,j}^{(I)}(\tau) = \frac{\overline{\Delta r_i^{(I)}(t+\tau) \Delta r_j^{(I)}(t)}}{\sqrt{\overline{\Delta {r_i^{(I)}}^2(t)}}\sqrt{\overline{\Delta {r_j^{(I)}}^2(t)}}},
\end{equation}
where $\Delta r_i^{(I)}(t) = r_i^{(I)}(t) - \overline{r_i^{(I)}(t)}$ and the overline denotes the time average.
Here, the number of data used for the calculation of each temporal cross-correlation function $C_{i,j}^{(I)}(\tau)$ is $4 \times 10^3$.
Similar to the case of non-modular small-world network \cite{Kim}, we introduce the spatial cross-correlation function $C_l^{(I)}$ ($l=1,...,L/2)$ between neuronal pairs separated by a spatial distance $l$ in the $I$th sub-network through average of all the temporal cross-correlations between $r_i^{(I)}(t)$ and $r_{i+l}^{(I)}(t)$ $(i=1,...,L)$ at the zero-time lag:
\begin{equation}
C_l^{(I)} = \frac{1}{L} \sum_{i=1}^{L} C_{i, i+l}^{(I)}(0) ~~~~ {\rm for~} l=1, \cdots, L/2.
\label{eq:SCC}
\end{equation}
Here, if $i+l > L$ in Eq.~(\ref{eq:SCC}), then $i+l-L$ is considered instead of $i+l$ because neurons lie on the ring.
If the spatial cross-correlation function $C_l^{(I)}$ ($l=1,...,L/2)$ is non-zero in the whole range of $l$, then the spatial correlation length $\eta_I$ becomes $L/2$ (note that the maximal distance between neurons is $L/2$ because of the ring architecture on which neurons exist) covering the whole sub-network. For this case, synchronization appears in the sub-network; otherwise, unsynchronization occurs.

\subsection{Cross-correlation Modularity Measure}
\label{subsec:CMM}
To determine the type of synchronization (modular or global) in modular networks, we measure the matching degree between the intra-modular dynamics of sub-networks in terms of the cross-correlation modularity measure $C_M$, based on ISPSRs. The normalized temporal cross-correlation function $C_{I,J}(\tau)$ between the ISPSR kernel estimates $R_s^{(I)}(t)$ and $R_s^{(J)}(t)$ of the $I$th and the $J$th sub-networks is given by:
\begin{equation}
C_{I,J}(\tau) = \frac{\overline{\Delta R_s^{(I)}(t+\tau) \Delta R_s^{(J)}(t)}}{\sqrt{\overline{\Delta {R_s^{(I)}}^2(t)}}\sqrt{\overline{\Delta {R_s^{(J)}}^2(t)}}},
\end{equation}
where $\Delta R_s^{(I)}(t) = R_s^{(I)}(t) - \overline{R_s^{(I)}(t)}$ and the overline denotes the time average. Figures \ref{fig:CMM}(c1)-\ref{fig:CMM}(c5) show the normalized temporal cross-correlation functions $C_{I,J}(\tau)$ for $J_{inter}=$10, 30, 70, 400, and 1200, respectively. Then, the cross-correlation modularity measure $C_M$ is obtained through average of the temporal cross-correlations between all the sub-population pairs at the zero-time lag:
\begin{equation}
C_M = \frac{2}{M (M-1)} \sum_{I=1}^{M-1} \sum_{J=I+1}^{M} C_{I,J}(0).
\label{eq:CMM}
\end{equation}
Here, the cross-correlation modularity measure $\langle C_M \rangle_r$ is obtained through average over 20 realizations, and the number of data used for the calculation of each temporal cross-correlation function $C_{I,J}(\tau)$ in each realization is $4 \times 10^3$.

\subsection{State Diagram}
\label{subsec:SD}
Population states vary depending on the inter-modular connection parameters $J_{inter}$ and $M_{syn}^{(inter)}$, which may be well shown in the state diagram in the $J_{inter}-M_{syn}^{(inter)}$ plane. To obtain the state diagram, we first divide the $J_{inter}-M_{syn}^{(inter)}$ plane into the $20 \times 10$ grids. Then, at each grid point, we calculate the sub-population order parameters ${\cal{O}}_s^{(I)}$ $(I=1, 2, 3)$ for $L=10^3$ and $10^4$ to determine whether the population state at the grid point is synchronized or unsynchronized. If ${\cal{O}}_s^{(I)}$ for $L=10^4$ is smaller than $f \cdot {\cal{O}}_s^{(I)}$ for $L=10^3$ ($f$ is some appropriate factor less than unity; for convenience we set $f=0.3$), ${\cal{O}}_s^{(I)}$ is expected to decrease with increasing $L$. For the case of decrease in ${\cal{O}}_s^{(I)}$ with increasing $L$, unsynchronization occurs at the grid point; otherwise, synchronization appears. Next, at the grid points where synchronization occurs, we calculate the  cross-correlation modularity measure $C_M$ to determine whether the population synchronization is modular or global. If $C_M$ is larger than a threshold $C_M^{th}$ (in our computation, we set $C_M^{th} =0.995$), global synchronization is expected to appear at the grid point; otherwise, modular synchronization occurs. After determining the population states (modular or global synchronization and unsynchronization) at all grid points, we try to obtain the synchronization-unsynchronization and the modular-global synchronization transition curves accurately. To this end, we calculate ${\cal{O}}_s^{(I)}$ $(C_M)$ in the small parameter region between the synchronization and unsynchronization (the modular and global synchronization) grid points by varying $J_{inter}$ or $M_{syn}^{(inter)}$. Moreover, to get more accurate transition curves, we divide a part of the parameter plane where the transition curves change rapidly into more minute grids and repeat the above computations.

\subsection{Sub- and Whole-population Statistical-Mechanical Spiking Measures}
\label{subsec:SWSM}
We measure the degree of modular and global sparse synchronization in terms of realistic statistical-mechanical sub- and whole-population spiking measures, based on the ISPSR and the IWPSR kernel estimates $R_s^{(I)}(t)$ and $R_w(t)$ \cite{RM}. Spike synchronization may be well visualized in the raster plots of spikes. For a synchronized case, the raster plot is composed of partially-occupied stripes (indicating sparse synchronization), and the corresponding ISPSR and IWPSR kernel estimates, $R_s^{(I)}(t)$ ($I=1,2,3$) and $R_w(t)$, exhibit regular oscillations. Each $i$th ($i=1,2,3,...$) global cycles of $R_s^{(I)}(t)$ and $R_w(t)$ begin from their left minimum, pass the central maximum, and end at the right minimum [also, corresponding to the beginning point of the next $(i+1)$th global cycles]; the 1st global cycles of $R_s^{(I)}(t)$ and $R_w(t)$ appear after transient times of $10^3$ ms, respectively. Spikes which appear in the $i$th global cycles of $R_s^{(I)}(t)$ and $R_w(t)$ form the $i$th stripes in the raster plots for the sub- and the whole-populations, respectively.
To measure the degree of the sub- and the whole-population spike synchronization seen in the raster plots, statistical-mechanical sub- and whole-population measures $M_s^{(I)}$ and $M_s^{(w)}$, based on $R_s^{(I)}(t)$ and $R_w(t)$, are introduced by considering the occupation pattern and the pacing pattern of spikes in the stripes for the sub- and the whole-populations, which corresponds to a simple extension of the case of non-modular networks \cite{RM}. The sub- and the whole-population spiking measures $M_i^{(I)}$ $(I=1$, 2, 3) and $M_i^{(w)}$ of the $i$th stripes [appearing in the $i$th global cycles of $R_s^{(I)}(t)$ and $R_w(t)$] are defined by the products of the sub- and the whole-population occupation degrees $O_i^{(I)}$ and $O_i^{(w)}$ of spikes (representing the density of the $i$th stripes) and the sub- and the whole-population pacing degrees $P_i^{(I)}$ and $P_i^{(w)}$ of spikes (denoting the smearing of the $i$th stripes), respectively:
\begin{equation}
  M_i^{(I)} = O_i^{(I)} \cdot P_i^{(I)}~~~{\rm and}~~~ M_i^{(w)} = O_i^{(w)} \cdot P_i^{(w)}.
\label{eq:SM}
\end{equation}
The sub- and the whole-population occupation degrees $O_i^{(I)}$ and  $O_i^{(w)}$ in the $i$th stripes are given by the fractions of spiking neurons in the $i$th stripes:
\begin{equation}
   O_i^{(I)} = \frac {N_{I,i}^{(s)}} {L} ~~~{\rm and}~~~ O_i^{(w)} = \frac {N_{w,i}^{(s)}} {M \cdot L}
\label{eq:Occu}
\end{equation}
where $N_{I,i}^{(s)}$ and $N_{w,i}^{(s)}$ are the numbers of spiking neurons in the $i$th stripes for the $I$th sub-network and the whole network, respectively.
For sparse synchronization with partially-occupied stripes, $O_i^{(I)}\ll 1$ and $O_i^{(w)}\ll 1$. The pacing degrees $P_i^{(I)}$ and $P_i^{(w)}$ of sparse spikes in the $i$th stripes for the sub- and the whole-populations can be determined in a statistical-mechanical way by taking into account their contributions to the macroscopic ISPSR and IWPSR kernel estimates $R_s^{(I)}(t)$ and $R_w(t)$, respectively. Instantaneous global phases $\Phi_s^{(I)}(t)$ of $R_s^{(I)}(t)$ and $\Phi_w(t)$ of $R_w(t)$ are introduced via linear interpolation in the two successive subregions forming global cycles \cite{RM}. The global phases $\Phi_s^{(I)}(t)$ and $\Phi_w(t)$ between the left minimum (corresponding to the beginning point of the $i$th global cycle) and the central maximum are given by
\begin{eqnarray}
\Phi_s^{(I)}(t) &=& 2\pi(i-3/2) + \pi \left(\frac{t-t_{I,i}^{(min)}}{t_{I,i}^{(max)}-t_{I,i}^{(min)}} \right) {\rm~~ for~} ~t_{I,i}^{(min)} \leq  t < t_{I,i}^{(max)}, \\
\Phi_w(t) &=& 2\pi(i-3/2) + \pi \left(\frac{t-t_{w,i}^{(min)}}{t_{w,i}^{(max)}-t_{w,i}^{(min)}} \right) {\rm~~ for~} ~t_{w,i}^{(min)} \leq  t < t_{w,i}^{(max)},
\end{eqnarray}
and $\Phi_s^{(I)}(t)$ and $\Phi_w(t)$ between the central maximum and the right minimum [corresponding to the beginning point of the $(i+1)$th global cycle]
are given by
\begin{eqnarray}
\Phi_s^{(I)}(t) &=& 2\pi(i-1) + \pi \left(\frac{t-t_{I,i}^{(max)}}{t_{I,i+1}^{(min)}-t_{I,i}^{(max)}} \right) {\rm~~ for~} ~t_{I,i}^{(max)} \leq  t < t_{I,i+1}^{(min)}, \\
\Phi_w(t) &=& 2\pi(i-1) + \pi \left(\frac{t-t_{w,i}^{(max)}}{t_{w,i+1}^{(min)}-t_{w,i}^{(max)}} \right) {\rm~~ for~} ~t_{w,i}^{(max)} \leq  t < t_{w,i+1}^{(min)},
\end{eqnarray}
where $t_{I,i}^{(min)}$ and $t_{w,i}^{(min)}$ are the beginning times of the $i$th ($i=1, 2, 3, \cdots$) global cycles of $R_s^{(I)}(t)$ and $R_w(t)$ [i.e., the times at which the left minima of $R_s^{(I)}(t)$ and $R_w(t)$ appear in the $i$th global cycles], respectively, and $t_{I,i}^{(max)}$ and $t_{w,i}^{(max)}$ are the times at which the maxima of $R_s^{(I)}(t)$ and $R_w(t)$ appear in the $i$th global cycles, respectively. Then, the contributions of the $k$th microscopic spikes in the $i$th stripes occurring at the times $t_{I,k}^{(s)}$ and $t_{w,k}^{(s)}$ to $R_s^{(I)}(t)$ and $R_w(t)$ are given by $\cos \Phi_k^{(I)}$ and $\cos \Phi_k^{(w)}$, where $\Phi_k^{(I)}$ and $\Phi_k^{(w)}$ are the global phases at the $k$th spiking times [i.e., $\Phi_k^{(I)} \equiv \Phi_s^{(I)}(t_{I,k}^{(s)})$ and $\Phi_k^{(w)} \equiv \Phi_w(t_{w,k}^{(s)})$]. Microscopic spikes make the most constructive (in-phase) contributions to $R_s^{(I)}(t)$ and $R_w(t)$ when the corresponding global phases $\Phi_k^{(I)}$ and $\Phi_k^{(w)}$ are $2 \pi n$ ($n=0,1,2, \dots$), while they make the most destructive (anti-phase) contribution to $R_s^{(I)}(t)$ and $R_w(t)$ when $\Phi_k^{(I)}$ and $\Phi_k^{(w)}$ are $2 \pi (n-1/2)$. By averaging the contributions of all microscopic spikes in the $i$th stripes to $R_s^{(I)}(t)$ and $R_w(t)$, we obtain the pacing degrees $P_i^{(I)}$ and $P_i^{(w)}$ of spikes in the $i$th stripes:
\begin{equation}
 P_i^{(I)} = { \frac {1} {S_i^{(I)}}} \sum_{k=1}^{S_i^{(I)}} \cos \Phi_k^{(I)}~~~{\rm and}~~~
 P_i^{(w)} = { \frac {1} {S_i^{(w)}}} \sum_{k=1}^{S_i^{(w)}} \cos \Phi_k^{(w)},
\label{eq:Pacing}
\end{equation}
where $S_i^{(I)}$ and $S_i^{(w)}$ are the total numbers of microscopic spikes in the $i$th stripes for the sub- and the whole-population, respectively. By averaging $M_i^{(I)}$ and $M_i^{(w)}$ of Eq.~(\ref{eq:SM}) over a sufficiently large number $N_s$ of stripes, we obtain the sub- and whole-population statistical-mechanical spiking measures $M_s^{(I)}$ and $M_s^{(w)}$:
\begin{equation}
M_s^{(I)} =  {\frac {1} {N_s}} \sum_{i=1}^{N_s} M_i^{(I)}~~~{\rm and}~~~
M_s^{(w)} =  {\frac {1} {N_s}} \sum_{i=1}^{N_s} M_i^{(w)}.
\label{eq:SM2}
\end{equation}
Here, we follow $3 \times 10^3$ global cycles in each realization, and obtain average occupation degrees, average pacing degrees, and average statistical-mechanical spiking measures via average over 20 realizations.

\section{Effect of Inter-Modular Connection on Fast Sparsely Synchronized Rhythms}
\label{sec:EIC}
In this section, we investigate the effect of inter-modular connection on emergence of fast sparsely synchronized rhythms by varying both the inter-modular coupling strength $J_{inter}$ and the average number of inter-modular links per interneuron $M_{syn}^{(inter)}$ in the clustered small-world network of inhibitory FS Izhikevich interneurons. In contrast to the case of non-modular networks, two kinds of sparsely synchronized states  such as modular and global sparse synchronization are thus found. These sparsely synchronized states are characterized by employing diverse realistic measures, explained in Sec.~\ref{sec:Method}.

In the absence of inter-modular coupling, we consider three cases of the intra-modular dynamics in the Watts-Strogatz small-world sub-networks: (1) synchronized in all identical sub-networks, (2) unsynchronized in all identical sub-networks, and (3) synchronized/unsynchronized in non-identical sub-networks. For each case, we study emergence of sparsely synchronized population states by changing both $J_{inter}$ and $M_{syn}^{(inter)}$
for a fixed set of $I_{DC}$, $D$, $M_{syn}^{(intra)}$, and $J_{syn}^{(intra)}$ (whose values are listed in Table \ref{tab:Parm}). In the subsection \ref{subsec:Case1}, we start from the 1st case of intra-modular dynamics. To further examine dependence on the type of intra-modular dynamics, we also study the 2nd and the 3rd cases in the subsection \ref{subsec:Case23}.

\subsection{1st Case of Intra-Modular Dynamics: Synchronized in All Identical Sub-networks}
\label{subsec:Case1}
\begin{figure}
\includegraphics[width=0.8\columnwidth]{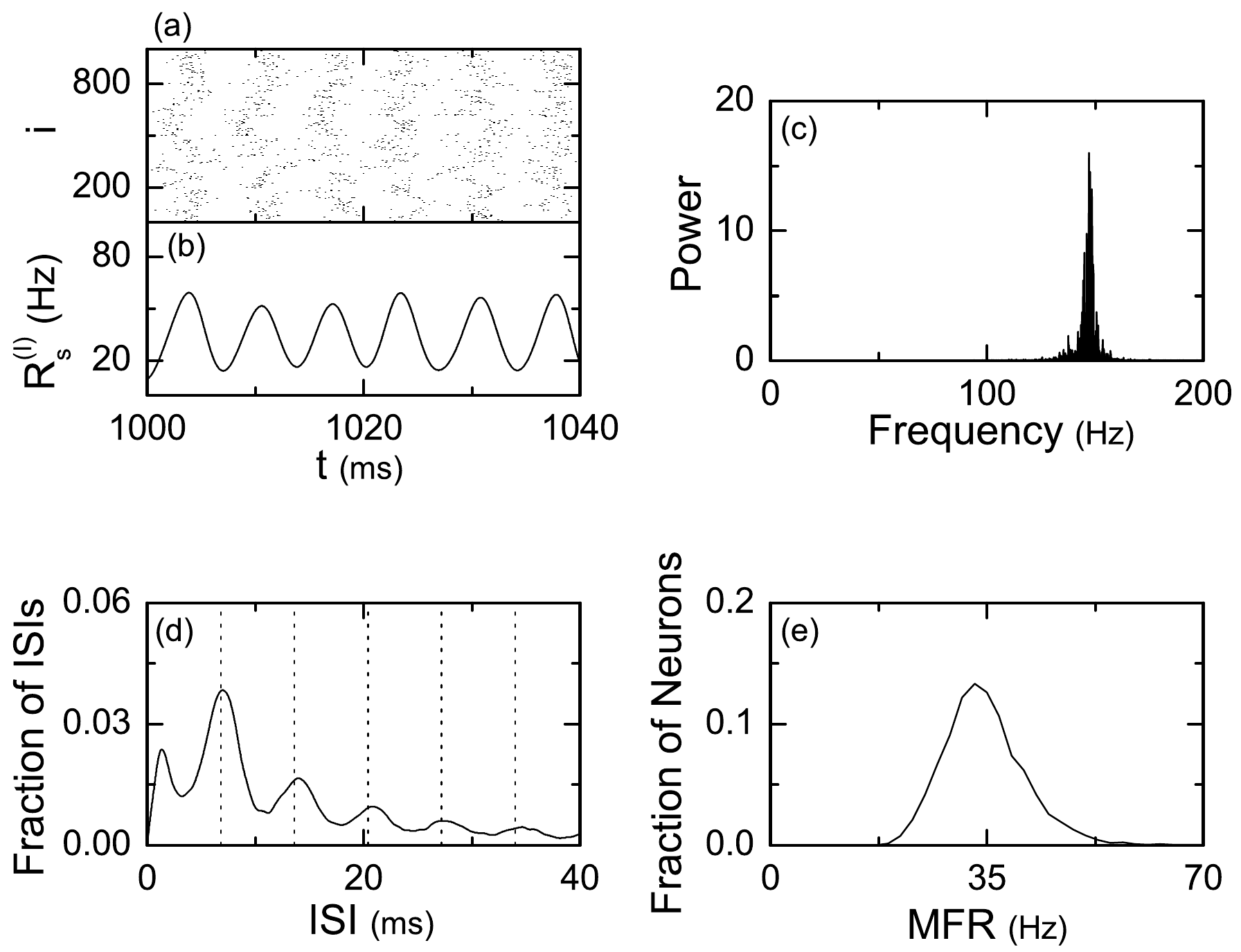}
\caption{Fast sparse synchronization in identical small-world sub-networks with $p_{rewiring}=0.25$ (case 1) in the absence of inter-modular coupling. (a) Raster plot of neural spikes and (b) instantaneous sub-population spike rate kernel estimate $R_s^{(I)}(t)$. (c) One-sided power spectrum of $\Delta R_s^{(I)}(t)$ $[= R_s^{(I)}(t) - \overline{R_s^{(I)}(t)}]$ with the mean-squared amplitude normalization. (d) Inter-spike interval (ISI) histogram for the individual interneurons; the vertical dotted lines denote integer multiples of the global period $T_I$ $(\simeq 6.8$ ms) of $R_s^{(I)}(t)$. (e) MFR (mean firing rate) distribution of individual interneurons.
}
\label{fig:SD}
\end{figure}

In this subsection, we consider the 1st case of intra-modular dynamics which are synchronized in 3 identical small-world sub-networks with the same rewiring probability $p_{rewiring}=0.25$.
In the absence of inter-modular coupling (i.e., $J_{inter}=0$), every sub-population in the small-world sub-networks exhibits identical sparse synchronization, as shown in Fig.~\ref{fig:SD}. Clear stripes are formed in the raster plot of Fig.~\ref{fig:SD}(a). The density of stripes is sparse because only a small fraction (about $0.22$) of the total $L$ $(=10^3)$ neurons in the sub-population fire in each stripe. Due to presence of these sparse stripes, the ISPSR kernel estimate $R_s^{(I)}(t)$ shows fast regular oscillation with sub-population frequency $f_p^{(I)} \simeq 147$ Hz, as shown in Figs.~\ref{fig:SD}(b)-\ref{fig:SD}(c). For the case of individual neurons, the inter-spike interval (ISI) histogram has multiple peaks appearing at multiples of the period $T_I$ ($\simeq 6.8$ ms) of $R_s^{(I)}(t)$ (i.e., skipping of spikes occurs at random integer multiples of $T_I$) [see Fig.~\ref{fig:SD}(d)]. Because of this stochastic spike skipping (also called the stochastic phase locking) \cite{Kim1,Kim2,Kim3,Kim,Kim-SFN,RM,GR,Longtin1,Longtin2}, individual neurons exhibit stochastic and intermittent spike discharges, and hence partial occupation occurs in the stripes of the raster plot. In contrast to sub-population rhythms, the distribution of mean firing rates (MFRs) of individual neurons shows a peak near $f_i^{(I)}$ $(\simeq 33$ Hz) which is much less than the sub-population frequency $f_p^{(I)}$ [see Fig.~\ref{fig:SD}(e)]. In this way, firing activity of individual neurons differs distinctly from the population oscillatory behavior for the case of sparse synchronization \cite{W_Review,Sparse1,Sparse2,Sparse3,Sparse4,Sparse5,Sparse6,Kim,Kim-SFN,RM}. For more details on the sparse synchronization in the (non-modular) small-world network, refer to \cite{Kim}.

From now on, we employ the methods for characterizing population dynamics in Sec.~\ref{sec:Method}, and investigate the effect of inter-modular connection on sparse synchronization by changing the inter-modular coupling strength $J_{inter}$ for $M_{syn}^{(inter)}=20$. Figures \ref{fig:SD1}(a)-\ref{fig:SD1}(c) show the raster plots of spikes in the three sub-populations for $J_{inter}=10$, 500, and 2500, respectively. The corresponding ISPSR and IWPSR kernel estimates, $R_s^{(I)}(t)$ and $R_w(t)$ of Eqs.~(\ref{eq:ISPSR}) and (\ref{eq:IWPSR}), for $J_{inter}=10$, 500, and 2500 are also shown in Figs.~\ref{fig:SD1}(d1)-\ref{fig:SD1}(f4), respectively.
For small $J_{inter}$, the inter-modular coupling strength plays a constructive role to favor the pacing between spikes in each sub-network, as shown in the case of $J_{inter}=10$. For each $I$th sub-population, sparse stripes are formed in the raster plot and $R_s^{(I)}(t)$ shows a regular oscillation, as shown in Fig.~\ref{fig:SD1}(a) and Figs.~\ref{fig:SD1}(d1)-\ref{fig:SD1}(d4). Hence, each sub-population exhibits sparse synchronization. However, the intra-modular dynamics of sub-networks make some mismatching because both the stripes and the ISPSRs between the sub-networks are shifted. Vertical gray lines which pass minima of $R_w(t)$ are drawn as reference lines for matching between $R_s^{(I)}(t)$ $(I=1,2,3)$ [where the minima of $R_s^{(I)}(t)$ [$R_w(t)$] are denoted by solid (open) circles]. As a result of mismatching, the degree of sparse synchronization in the whole population becomes less than that in the sub-networks [i.e., the amplitude of $R_w(t)$ is less than that of $R_s^{(I)}(t)$], and this kind of population behavior for $J_{inter}=10$ is referred to as the modular sparse synchronization because it reveals the modular structure. With increasing $J_{inter}$, the mismatching degree between the intra-modular dynamics of sub-networks decreases, although the stripes in the raster plots become more sparse due to increased inhibition. Eventually when passing a threshold $J_{inter}^*$ $(\simeq 268)$, intra-modular dynamics of sub-networks begin to make perfect matching. As a result, the population behavior becomes globally identical, independently of the cluster structure, as shown in Fig.~\ref{fig:SD1}(b) and Figs.~\ref{fig:SD1}(e1)-\ref{fig:SD1}(e4) for $J_{inter}=500$ [where all the minima of $R_s^{(I)}(t) (I=1,2,3)$ lie on the reference vertical line passing the minima of $R_w$(t)], and it is referred to as the global sparse synchronization. However, for sufficiently large $J_{inter}$, due to strong inhibition the inter-modular coupling strength plays a destructive role to spoil the pacing between sparse spikes. Hence, as $J_{inter}$ passes a higher critical value $J_{inter,h}^*$ $(\simeq 1657)$ the global sparse synchronization breaks into unsynchronization. As an example, refer to the case of $J_{inter}=2500$. Sparse spikes in the raster plot of each sub-network are completely scattered without forming any stripes [see Fig.~\ref{fig:SD1}(c)], and hence each ISPSR kernel estimate $R_s^{(I)}(t)$ becomes nearly stationary (i.e., every sub-network exhibits an unsynchronized state), as shown in Figs.~\ref{fig:SD1}(f1)-\ref{fig:SD1}(f4). We now vary not only $J_{inter}$ but also $M_{syn}^{(inter)}$, and investigate emergence of modular and global sparse synchronization in the whole $J_{inter}$-$M_{syn}^{(inter)}$ plane by using the method explained in the subsection \ref{subsec:SD}. Thus, we obtain the state diagram in Fig.~\ref{fig:SD1}(g). Modular sparse synchronization emerges for small $J_{inter}$ or $M_{syn}^{(inter)}$ in the ``L''-shaped gray region, while in the dark gray region global sparse synchronization appears. For large $J_{inter}$ $(> 1572)$, unsynchronization occurs between the modular and the global synchronization. Changes in the population behaviors along the routes I, II, and III in Fig.~\ref{fig:SD1}(g) are given in the following subsubsections.

\subsubsection{Effect of The Inter-Modular Coupling Strength on Population Synchronization along The Route I}
\label{subsubsec:Route1}
\begin{figure}
\includegraphics[width=0.7\columnwidth]{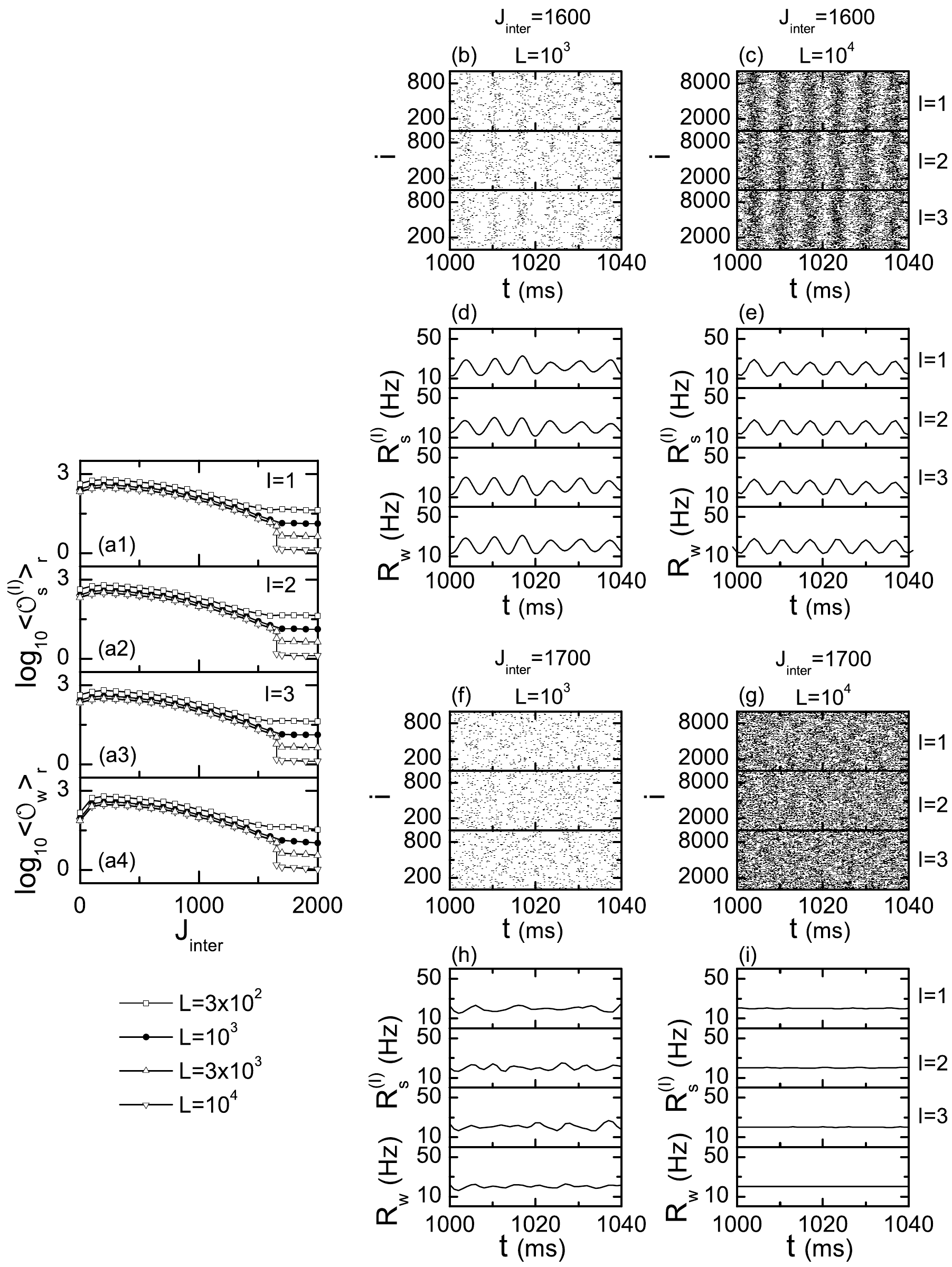}
\caption{Realistic thermodynamic order parameters for measurement of the threshold for the synchronization-unsynchronization transition along the route I with $M_{syn}^{(inter)}=20$ for the 1st case of intra-modular dynamics with $p_{rewiring}=0.25$. (a1)-(a4) Plots of the sub- and the whole-population order parameters $\log_{10}{\cal{O}}_s^{(I)}$ ($I=1,2,3$) and $\log_{10}{\cal{O}}_w$ versus $J_{inter}$. Sparse synchronization for $J_{inter}=1600$: raster plots of neural spikes in the three sub-networks for (b) $L=10^3$ and (c) $L=10^4$ and instantaneous sub- and whole-population spike rate kernel estimates $R_s^{(I)}(t)$ $(I=1,2,3)$ and $R_w(t)$ for (d) $L=10^3$ and (e) $L=10^4$. Unsynchronization for $J_{inter}=1700$: raster plots of neural spikes in the three sub-networks for (f) $L=10^3$ and (g) $L=10^4$ and instantaneous sub- and whole-population spike rate kernel estimates $R_s^{(I)}(t)$ $(I=1,2,3)$ and $R_w(t)$ for (h) $L=10^3$ and (i) $L=10^4$.
}
\label{fig:Order}
\end{figure}

In order to study the effect of the inter-modular coupling $J_{inter}$ on the population synchronization, we consider the case of the route I with $M_{syn}^{(inter)}=20$. Some results for this case are given for $J_{inter}=10$, 500, and 2500 in Figs.~\ref{fig:SD1}(a)-\ref{fig:SD1}(f4). As $J_{inter}$ is increased, a transition from modular sparse synchronization to global sparse synchronization when passing a threshold $J_{inter}^* (\simeq 268)$, and eventually to unsynchronization when passing a higher threshold $J_{inter,h}^* (\simeq 1657)$ occurs. The higher threshold $J_{inter,h}^*$ for the transition to unsynchronization is determined through calculation of the sub- and the whole-population order parameters $\langle {\cal{O}}_s^{(I)} \rangle_r$ and $\langle {\cal{O}}_w \rangle_r$ of Eqs.~(\ref{eq:SOrder}) and (\ref{eq:WOrder}), where $\langle \cdots \rangle_r$ denotes an average over realizations. Figures \ref{fig:Order}(a1)-\ref{fig:Order}(a4) show plots of $\langle {\cal{O}}_s^{(I)} \rangle_r$ and $\langle {\cal{O}}_w \rangle_r$ versus $J_{inter}$. For $J_{inter} < J_{inter,h}^*$ $(\simeq 1657$), synchronized states exist because the values of $\langle {\cal{O}}_s^{(I)} \rangle_r$ and $\langle {\cal{O}}_w \rangle_r$ become saturated to non-zero limit values for large $L$. When passing the higher threshold threshold $J_{inter,h}^*$, a transition to unsynchronization occurs because the order parameters $\langle {\cal{O}}_s^{(I)} \rangle_r$ and $\langle {\cal{O}}_w \rangle_r$ tend to zero as $L \rightarrow \infty$. These unsynchronized states seem to appear due to a destructive effect of strong inhibition spoiling the pacing between sparse spikes. Here, we present two explicit examples for the synchronized and the unsynchronized states. First, we consider the synchronized case for $J_{inter}=1600$. For $L=10^3$, sparse stripes are formed in the raster plot of spikes for each sub-network, and the ISPSR and the IWPSR kernel estimates $R_s^{(I)}(t)$ and $R_w(t)$ show regular oscillations, although there are some variations in the amplitudes [see Figs.~\ref{fig:Order}(b) and \ref{fig:Order}(d)]. As $L$ is increased to $L=10^4$, stripes in the raster plots become more clear, and $R_s^{(I)}(t)$ and $R_w(t)$ display more regular oscillations with nearly the same amplitudes, as shown in Figs.~\ref{fig:Order}(c) and \ref{fig:Order}(e). Consequently, the population state for $J_{inter}=1600$ seems to be synchronized because $R_s^{(I)}(t)$ and $R_w(t)$ tend to show more regular oscillations as $L$ goes to the infinity. As a second example, we consider the unsynchronized case of $J_{inter}=1700$. As shown in Fig.~\ref{fig:Order}(f) for $L=10^3$, sparse spikes are scattered without forming any stripes in the raster plot, and $R_s^{(I)}(t)$ and $R_w(t)$ in Fig.~\ref{fig:Order}(h) show little noisy fluctuations. In contrast to the synchronized case, as $L$ is increased to $L=10^4$, sparse spikes become more scattered, and consequently $R_s^{(I)}(t)$ and $R_w(t)$ become nearly stationary, as shown in Figs.~\ref{fig:Order}(g) and \ref{fig:Order}(i). Hence, the population state for $J_{inter}=1700$ seems to be unsynchronized because $R_s^{(I)}(t)$ and $R_w(t)$ tend to be nearly stationary as $L$ increases to the infinity.

\begin{figure}
\includegraphics[width=0.8\columnwidth]{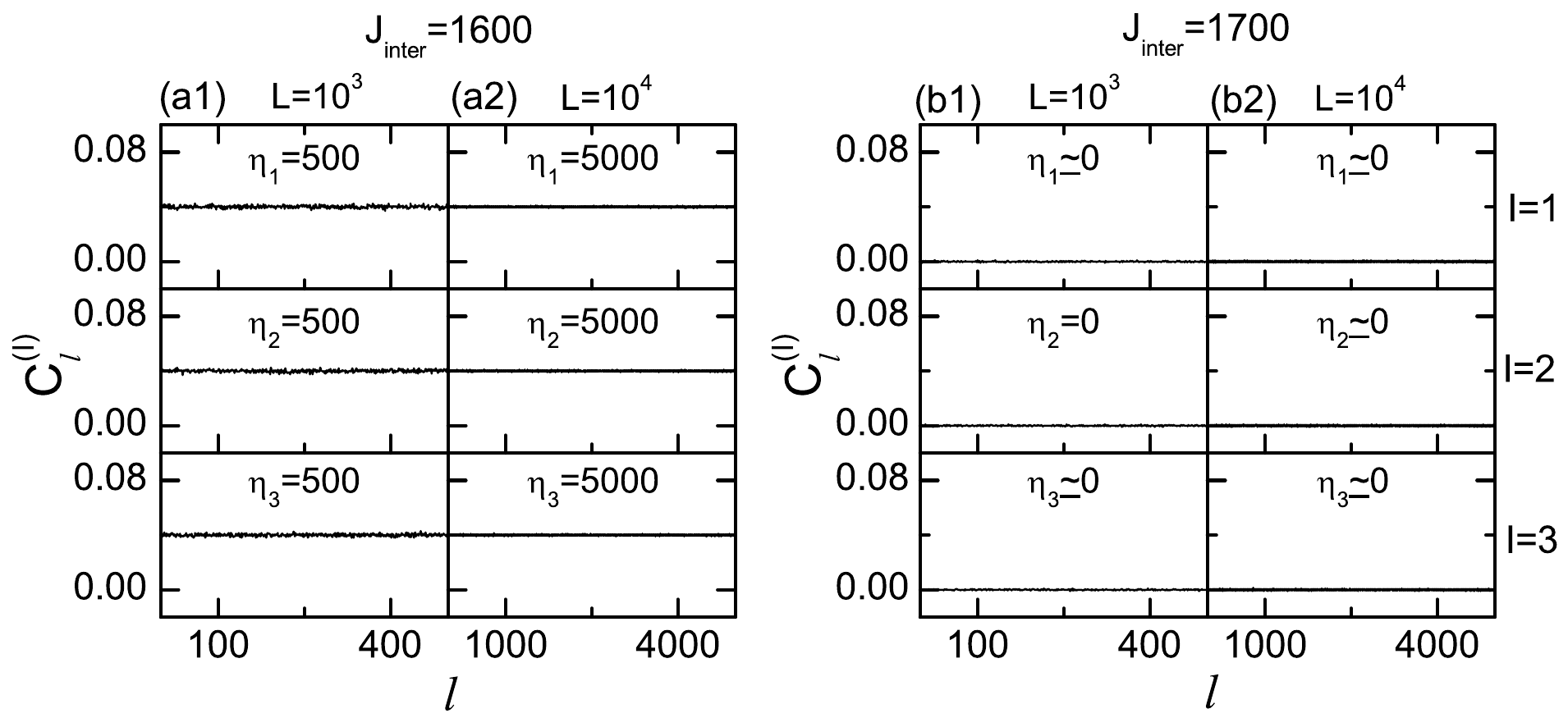}
\caption{Spatial cross-correlation functions for the synchronized and the unsynchronized states along the route I with $M_{syn}^{(inter)}=20$ for the 1st case of intra-modular dynamics with $p_{rewiring}=0.25$.
Sparse synchronization for $J_{inter}=1600$ with the spatial correlation length $\eta_I$ covering the whole sub-network: spatial cross-correlation functions $C_l^{(I)}$ in the three sub-networks ($I=1,2,3)$ for (a1) $L=10^3$ and (a2) $L=10^4$. Unsynchronization for $J_{inter}=1700$ with $\eta_I \simeq 0$: spatial cross-correlation functions $C_l^{(I)}$ in the three sub-networks for (b1) $L=10^3$ and (b2) $L=10^4$.
}
\label{fig:Corr}
\end{figure}

In order to further understand the above synchronization-unsynchronization transition, we investigate the effect of inter-modular connection on the ``microscopic'' dynamical cross-correlations between neuronal pairs. As examples, we reconsider the same cases of $J_{inter}=1600$ and 1700 as in Fig.~\ref{fig:Order}. Figure \ref{fig:Corr}(a1) shows the plots of the spatial cross-correlation functions $C_l^{(I)}$ of Eq.~\ref{eq:SCC} versus $l$ for $L=10^3$ in the case of $J_{inter}=1600$. These spatial correlation functions $C_l^{(I)}$ are nearly non-zero constant $(\simeq 0.04)$ in the whole range of $l$, and hence the correlation length $\eta_I$ becomes $L/2$ (=500) covering the whole sub-networks (note that the maximal distance between neurons is $L/2$ because of the ring architecture on which neurons exist). Consequently, each sub-network is composed of just one single synchronized block. For $L=10^4$, the flatness of $C_l^{(I)}$ in Fig.~\ref{fig:Corr}(a2) also extends to the whole range ($l=L/2=5000$) of the $I$th sub-network, and hence the correlation length becomes $\eta_I=5000$, which also covers the whole sub-network. In this way, for $J_{inter}=1600$, due to a constructive role of $J_{inter}$ favoring the pacing between sparse spikes, the correlation length $\eta_I$ seems to cover the whole sub-network, independently of $L$. For this case, the normalized correlation length $\tilde{\eta_I}$ ($= \frac {\eta_I} {L}$), representing the ratio of the correlation length $\eta_I$ to the sub-network size $L$ (i.e., denoting the relative size of synchronized blocks when compared to the whole sub-network size), has a non-zero limit value, $1/2$, and consequently synchronization emerges in each sub-network. In contrast, for $J_{inter}=1700$ the spatial cross-correlation functions $C_l^{(I)}$ are nearly zero, independently of $L$, as shown in Figs.~\ref{fig:Corr}(b1)-\ref{fig:Corr}(b2). For this case, due to a destructive role of $J_{inter}$ spoiling the pacing between sparse spikes, the correlation length $\eta_I$ becomes nearly zero, and hence no synchronization occurs in each sub-network.

\begin{figure}
\includegraphics[width=0.7\columnwidth]{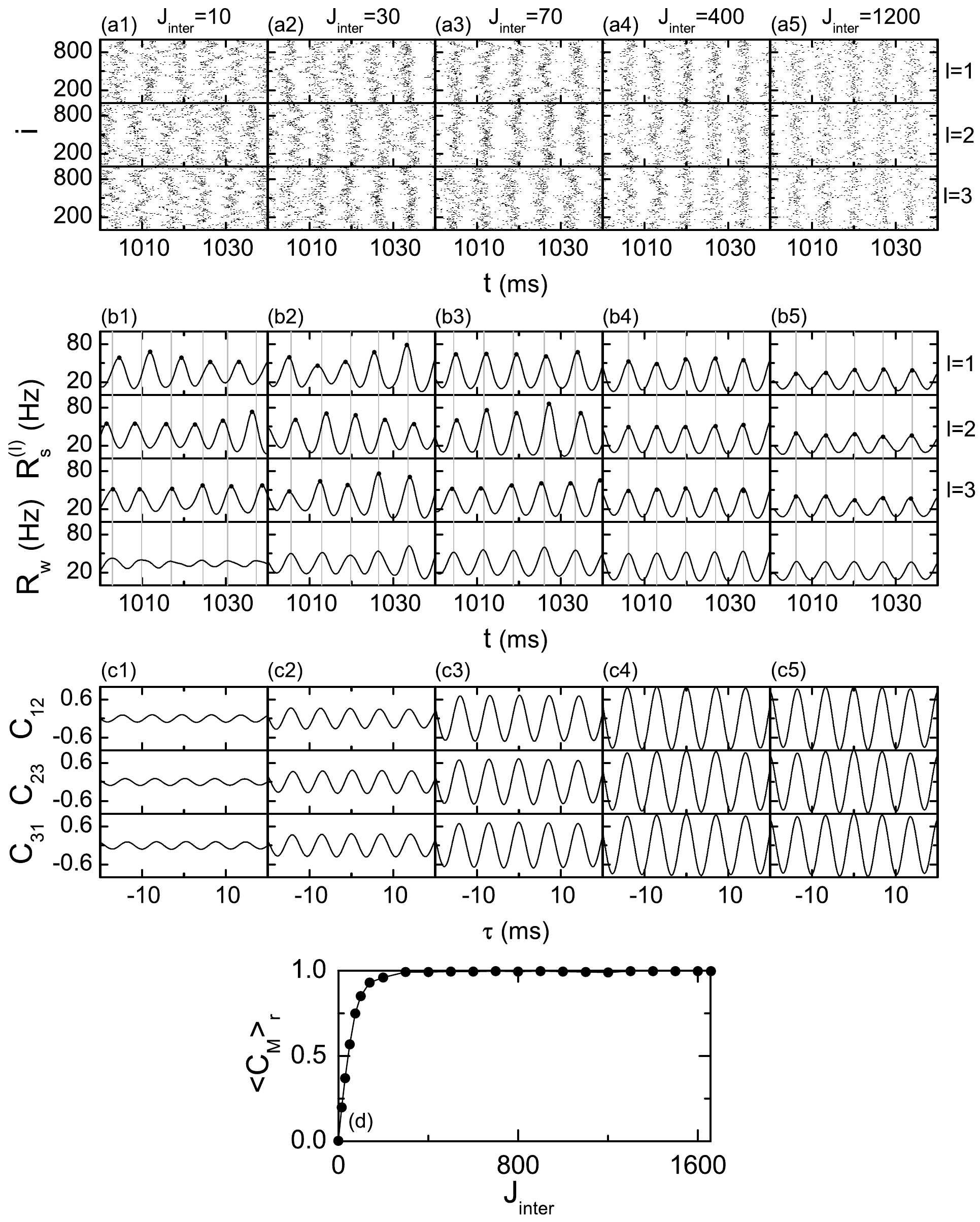}
\caption{Realistic cross-correlation modularity measure for determining the threshold for modular-global sparse synchronization transition along the route I with $M_{syn}^{(inter)}=20$ for the 1st case of intra-modular dynamics with $p_{rewiring}=0.25$. Raster plots of neural spikes in the sub-networks ($I=1,2,3)$ for $J_{inter}=$ (a1) 10, (a2) 30, (a3) 70, (a4) 400 and (a5) 1200. Instantaneous sub- and whole-population spike rate kernel estimates $R_s^{(I)}$ $(I=1,2,3)$ and $R_w(t)$ for $J_{inter}=$ (b1) 10, (b2) 30, (b3) 70, (b4) 400 and (b5) 1200. Vertical gray lines pass minima of $R_w(t)$, and the minima of $R_s^{(I)}(t)$ [$R_w(t)$] are represented by solid (open) circles. Temporal cross-correlation functions $C_{I,J}(\tau)$ between the instantaneous sub-population spike rate kernel estimates $R_s^{(I)}(t)$ and $R_s^{(J)}(t)$ of the sub-networks $I$ and $J$ for $J_{inter}=$ (c1) 10, (c2) 30, (c3) 70, (c4) 400 and (c5) 1200. (d) Plot of the cross-correlation modularity measure $\langle C_M \rangle_r$ versus $J_{inter}$.
}
\label{fig:CMM}
\end{figure}

We now investigate the type of synchronization through measurement of the matching degree between the intra-modular dynamics in sub-networks in the synchronized range of $0 < J_{inter} < J_{inter,h}^* (\simeq 1657)$ along the route I in Fig.~\ref{fig:SD1}(g). Figures \ref{fig:CMM}(a1)-\ref{fig:CMM}(a5) show the raster plots of spikes in the three sub-networks for $J_{inter}=$10, 30, 70, 400, and 1200, respectively. The ISPSR and the IWPSR kernel estimates, $R_s^{(I)}(t)$ and $R_w(t)$, for $J_{inter}=$10, 30, 70, 400, and 1200 are also shown in Figs.~\ref{fig:CMM}(b1)-\ref{fig:CMM}(b5), respectively. For each $I$th sub-population, sparse stripes are formed in the raster plot of spikes and the ISPSR kernel estimate $R_s^{(I)}(t)$ shows a regular oscillation with global frequency $f_p^{(I)} \simeq 147$ Hz. Hence, each sub-population shows sparse synchronization. For the case of modular sparse synchronization for $J_{inter}=10$, 30, and 70, the intra-modular dynamics of sub-networks make some mismatching because both the stripes and the ISPSR kernel estimates between the sub-networks are shifted [see Figs.~\ref{fig:CMM}(b1)-\ref{fig:CMM}(b3) where the minima of $R_s^{(I)}(t)$ (denoted by solid circles) lie off the reference vertical lines which pass the minima of $R_w(t)$ (represented by open circles)]. Hence the amplitude of $R_w(t)$ becomes less than that of $R_s^{(I)}(t)$. As $J_{inter}$ is increased, the mismatching degree decreases, and hence the amplitude of $R_w(t)$ increases. Eventually, when passing a threshold $J_{inter}^*$ $(\simeq 268)$ global sparse synchronization occurs. Hence, for $J_{inter}=400$ and 1200, intra-modular dynamics of sub-networks (shown in their raster plots and ISPSR kernel estimates) make perfect matching [i.e., the minima of $R_s^{(I)}(t)$ lie on the reference vertical lines, as shown in Figs.~\ref{fig:CMM}(b4)-\ref{fig:CMM}(b5)], and hence the amplitude of $R_w(t)$ becomes the same as that of $R_s^{(I)}(t)$. The matching degree between the intra-modular dynamics of subnetworks may be measured through calculation of the cross-correlation modularity measure $C_M$ of Eq.~\ref{eq:CMM}. Figure \ref{fig:CMM}(d) shows the plot of $\langle C_M \rangle_r$ versus $J_{inter}$ where $\langle \cdots \rangle_r$ denotes average over realizations. As $J_{inter}$ is increased, $\langle C_M \rangle_r$ increases monotonically, and eventually when passing the threshold $J_{inter}^*$ $(\simeq 268)$, its value becomes 1. Hence, for $J_{inter} < J_{inter}^*$ modular sparse synchronization (with $\langle C_M \rangle_r <1)$ emerges, while global sparse synchronization (with $\langle C_M \rangle_r=1$) appears for $J_{inter}^* < J_{inter} < J_{inter,h}^*$.

We also measure the degree of modular and global sparse synchronization in the synchronized range of $0 < J_{inter} < J_{inter,h}^*$. As shown in Figs.~\ref{fig:CMM}(a1)-\ref{fig:CMM}(a5), spike synchronization may be well visualized in the raster plots of spikes. For a synchronous case, ``stripes'' (composed of spikes and representing population synchronization) appear successively in the raster plot. For measurement of the degree the sub- and the whole-population spike synchronization seen in the raster plots, realistic statistical-mechanical sub- and whole-population measures $M_s^{(I)}$ and $M_s^{(w)}$ are introduced by considering the occupation pattern (representing the density of the stripes) and the pacing pattern (denoting the smearing of the stripes) of spikes in the stripes for the sub- and the whole-populations, as explained in the subsection \ref{subsec:SWSM}. By varying $J_{inter}$, we follow $3 \times 10^3$ stripes (i.e., $3 \times 10^3$ global cycles) in each realization, and through an average over 20 realizations, we obtain the sub- and the whole-population occupation degrees $\langle O_s^{(I)} \rangle_r$ and $\langle O_w \rangle_r$ of Eq.~(\ref{eq:Occu}), the sub- and the whole-population pacing degrees $\langle P_s^{(I)} \rangle_r$ and $\langle P_w \rangle_r$ of Eq.~(\ref{eq:Pacing}), and the statistical-mechanical sub- and whole-population spiking measures $M_s^{(I)}$ and $M_s^{(w)}$ of Eq.~(\ref{eq:SM2}), and the results are shown in Figs.~\ref{fig:SM}(a1)-\ref{fig:SM}(c4). For the case of modular synchronization [occurring on the left region of the vertical dotted threshold line for $J_{inter}=J_{inter}^*$ ($\simeq 268$)], both the occupation degree $\langle O_w \rangle_r$ and the pacing degree $\langle P_w \rangle_r$ for the whole-population are less than those for the sub-populations because of mismatching between the intra-modular dynamics of  sub-networks. As $J_{inter}$ is increased, their mismatching degrees become smaller, and eventually $\langle O_w \rangle_r$ and $\langle P_w \rangle_r$ for the whole population become the same as those for the sub-populations for the case of global synchronization (occurring on the right region of the vertical dotted threshold line) due to perfect matching between the intra-modular dynamics of sub-networks. We first consider the occupation degree which characterizes the sparseness degree of population synchronization. For the sub-populations, the occupation degrees $\langle O_s^{(I)} \rangle_r$ decrease monotonically because of increase in inhibition with increasing $J_{inter}$. In the case of modular synchronization, typical IWPSR kernel estimates $R_w(t)$ show faster and smaller-amplitude oscillations with the whole-population frequency $f_p^{(w)}$ larger than the sub-population frequency $f_p^{(I)}$, and hence the occupation degree $\langle O_w \rangle_r$ for the whole-population becomes less than $\langle O_s^{(I)} \rangle_r$. As $J_{inter}$ is increased, $\langle O_w \rangle_r$ increases and approaches $\langle O_s^{(I)} \rangle_r$ due to decrease in the mismatching degree between the intra-modular dynamics of sub-networks, and eventually when passing the threshold $J_{inter}^*$ (i.e., in the case of global synchronization) they become the same and then decrease with increasing $J_{inter}$. We note that modular and global synchronization is sparse one because $\langle O_s^{(I)} \rangle_r$ is much less than unity [i.e., only a small fraction of the total $L$ $(=10^3)$ neurons in the sub-population fire in each stripe]. Next, we consider the pacing degree between  spikes in the stripes. For relatively small $J_{inter}$, with increasing $J_{inter}$ the sub-population pacing degree $\langle P_s^{(I)} \rangle_r$ increases due to a constructive role of $J_{inter}$ favoring the pacing between spikes, while for large $J_{inter}$ $\langle P_s^{(I)} \rangle_r$ decreases as $J_{inter}$ is increased because of a destructive role of $J_{inter}$ spoiling the pacing between spikes. Through competition between these constructive and destructive roles of $J_{inter}$  a ``plateau'' with high pacing degree is formed in a relatively wide region of intermediate $J_{inter}$ for the case of global sparse synchronization. The whole-population pacing degree $\langle P_w \rangle_r$ (which is less than or equal to $\langle P_s^{(I)} \rangle_r$) also exhibits similar behavior. Consequently, both the sub- and the whole-population statistical-mechanical spiking measures $M_s^{(I)}$ and $M_s^{(w)}$ (which are obtained by taking into consideration both the occupation and the pacing degrees of spikes in the stripes) show bell-shaped curves with their peaks at $J_{inter} \simeq$ 202 (corresponding to modular sparse synchronization) and 287 (corresponding to global sparse synchronization), respectively. For further understanding of the pacing degree between spikes, we also consider the spatial cross-correlations between neuronal pairs. Figures \ref{fig:SM}(d1)-\ref{fig:SM}(d5) show the spatial cross-correlation functions $C_l^{(I)}$ of Eq.~(\ref{eq:SCC}) for $J_{inter}=$10, 30, 70, 400, and 1200, respectively. For the case of relatively small $J_{inter}$, with increasing $J_{inter}$ the value of $C_l^{(I)}$ increases, but it decreases for large $J_{inter}$. For quantitative analysis, we introduce the sub-population spatial cross-correlation degree $\langle \langle C_l^{(I)} \rangle_l \rangle_r$ given by double averaging of the spatial cross-correlation function $C_l^{(I)}$ over all lengths $l$ and realizations. This sub-population spatial cross-correlation degree $\langle \langle C_l^{(I)} \rangle_l \rangle_r$ is a microscopic measure quantifying the cross-correlation degree between the microscopic IISR kernel estimates $r_i^{(I)}(t)$ without any explicit relation to the macroscopic occupation and pacing patterns of spikes. Figure \ref{fig:SM}(e) shows plots of $\langle \langle C_l^{(I)} \rangle_l \rangle_r$ (obtained through average over 20 realizations) versus $J_{inter}$ for $I=$1, 2, and 3. Similar to the case of the sub-population pacing degree $\langle P_s^{(I)} \rangle_r$, $\langle \langle C_l^{(I)} \rangle_l \rangle_r$ also display similar bell-shaped curves with peaks in the region of global synchronization. Hence, the statistical-mechanical pacing degree between spikes seems to be somewhat associated with the microscopic spatial cross-correlation degree between neuronal pairs.

\subsubsection{Effect of The Average Number of Inter-Modular Connections along the Routes II and III}
\label{subsubsec:Route23}
\begin{figure}
\includegraphics[width=0.8\columnwidth]{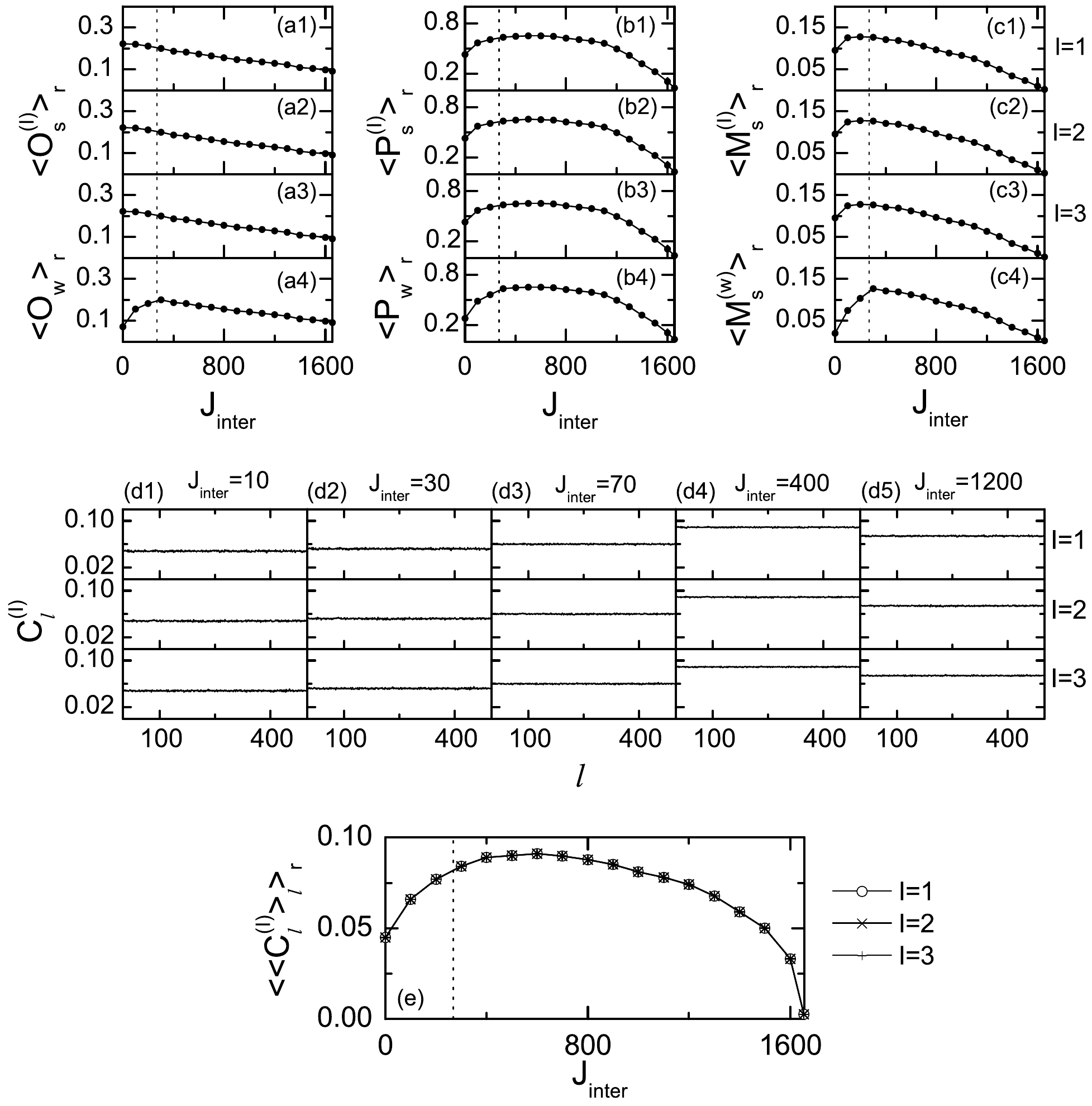}
\caption{Realistic statistical-mechanical spiking measure for measurement of the degree of modular and global sparse synchronization along the route I with $M_{syn}^{(inter)}=20$ for the 1st case of intra-modular dynamics with $p_{rewiring}=0.25$. Vertical dotted lines in (a1)-(c4) and (e) represent the threshold of $J=J_{inter}^*$ ($\simeq 268$). (a1)-(a4) Plots of the sub- and the whole-population occupation degrees  ${\langle O_s^{(I)} \rangle}_r$ and  ${\langle O_w \rangle}_r$ versus $J_{inter}$. (b1)-(b4) Plots of the sub- and the whole-population pacing degrees ${\langle P_s^{(I)} \rangle}_r$ and  ${\langle P_w \rangle}_r$ versus $J_{inter}$. (c1)-(c4) Plots of the sub- and the whole-population statistical-mechanical spiking measures ${\langle M_s^{(I)} \rangle}_r$ and  ${\langle M_s^{(w)} \rangle}_r$ versus $J_{inter}$. Spatial cross-correlation functions $C_l^{(I)}$ in the three sub-networks
for $J_{inter}=$ (d1) 10, (d2) 30, (d3) 70, (d4) 400 and (d5) 1200. (e) Plot of the spatial cross-correlation degree ${\langle {\langle C_l^{(I)} \rangle}_l \rangle}_r$ versus $J_{inter}$.
}
\label{fig:SM}
\end{figure}

\begin{figure}
\includegraphics[width=0.9\columnwidth]{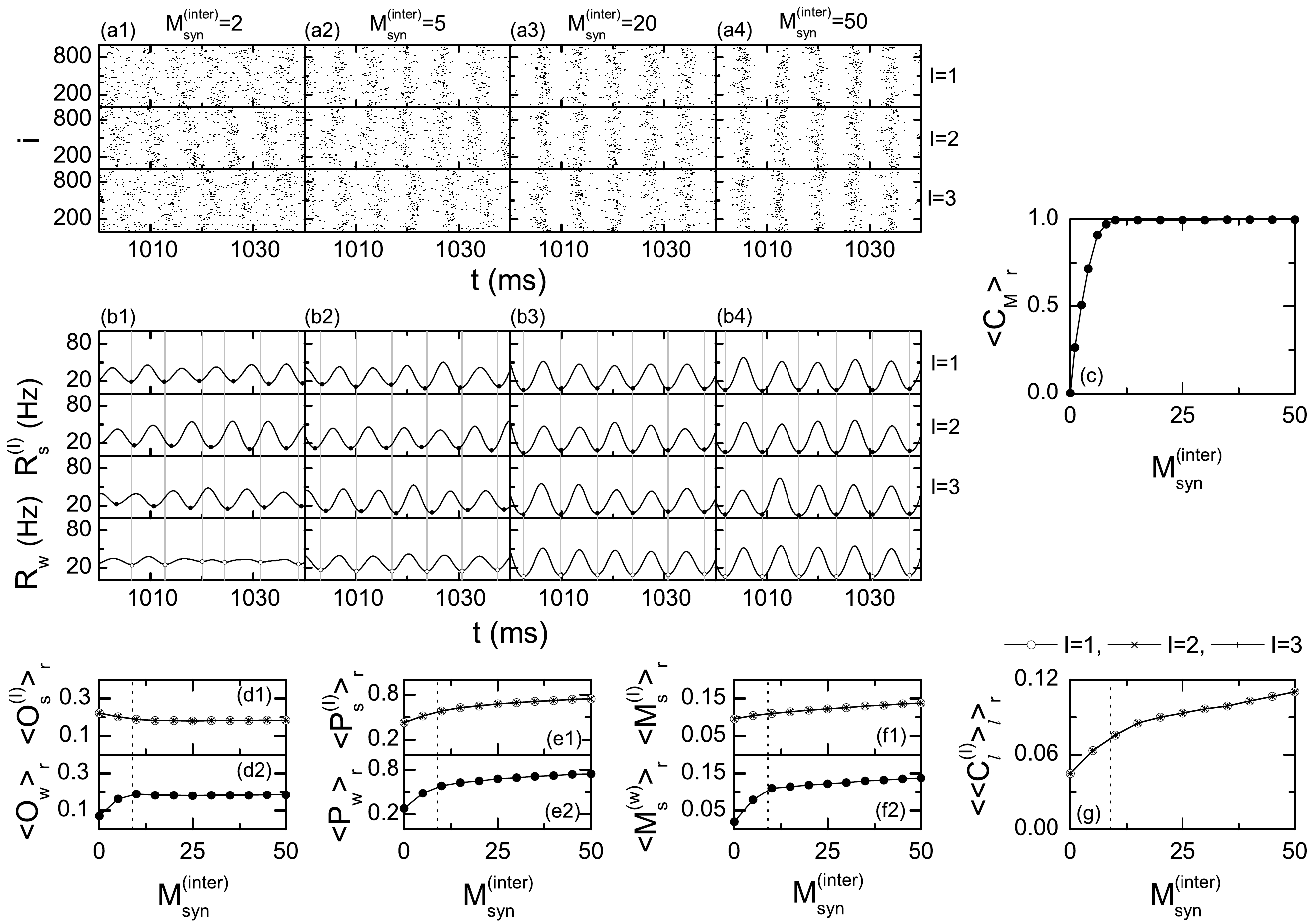}
\caption{Modular and global sparse synchronization along the route II with $J_{inter}=500$ for the 1st case of intra-modular dynamics with $p_{rewiring}=0.25$. Raster plots of neural spikes in the sub-networks ($I=1,2,3)$
for $M_{syn}^{(inter)}=$ (a1), (a2) 5, (a3) 20 and (a4) 50. Instantaneous sub- and whole-population spike rate kernel estimates $R_s^{(I)}(t)$ $(I=1,2,3)$ and $R_w(t)$ for $M_{syn}^{(inter)}=$ (a1), (a2) 5, (a3) 20 and (a4) 50.
Vertical gray lines pass  minima of $R_w(t)$, and the minima of $R_s^{(I)}(t)$ [$R_w(t)$] are represented by solid (open) circles. (c) Plot of the cross-correlation modularity measure $\langle C_M \rangle_r$ versus $M_{syn}^{(inter)}$. Vertical dotted lines in (d1)-(g) denote the threshold ${M_{syn}^{(inter)}}^* (\simeq 9)$. (d1)-(d2) Plots of the sub- and whole-population occupation degrees ${\langle O_s^{(I)} \rangle}_r$ and  ${\langle O_w \rangle}_r$ versus $M_{syn}^{(inter)}$. (e1)-(e2) Plots of the sub- and the whole-population pacing degrees  ${\langle P_s^{(I)} \rangle}_r$ and  ${\langle P_w \rangle}_r$ versus $M_{syn}^{(inter)}$. (f1)-(f2) Plots of the sub- and the whole-population statistical-mechanical spiking measures ${\langle M_s^{(I)} \rangle}_r$ and  ${\langle M_s^{(w)} \rangle}_r$ versus $M_{syn}^{(inter)}$. (g) Plot of the spatial cross-correlation degree ${\langle {\langle C_l^{(I)} \rangle}_l \rangle}_r$ versus $M_{syn}^{(inter)}$.
}
\label{fig:Route2}
\end{figure}

In addition to the above study on the effect of $J_{inter}$ along the route I for $M_{syn}^{(inter)}=20$, we also investigate the effect of average number of inter-modular connections per interneuron $M_{syn}^{(inter)}$ on emergence of modular and global sparse synchronization along the routes II and III for $J_{inter}=$ 500 and 2500, respectively [see Fig.~\ref{fig:SD1}(g)]. For the case of the route II with $J_{inter}=500$, the raster plots of spikes in the three sub-populations for $M_{syn}^{(inter)}=$ 2, 5, 20, and 50 are shown in Figs.~\ref{fig:Route2}(a1)-\ref{fig:Route2}(a4), respectively. The corresponding ISPSR and IWPSR kernel estimates, $R_s^{(I)}(t)$ and $R_w(t)$, for $M_{syn}^{(inter)}=$ 2, 5, 20, and 50 are also shown in Figs.~\ref{fig:Route2}(b1)-\ref{fig:Route2}(b4), respectively. For each $I$th sub-population, sparse stripes are formed in the raster plot and the ISPSR kernel estimate $R_s^{(I)}(t)$ shows a regular oscillation. As $M_{syn}^{(inter)}$ is increased, more clear stripes appear in the raster plots of sub-networks, and hence the amplitudes of $R_s^{(I)}(t)$ increase. Furthermore, with increasing $M_{syn}^{(inter)}$, the mismatching degree between the intra-modular dynamics of sub-networks decreases, and eventually when passing a threshold ${M_{syn}^{(inter)}}^*$ $(\simeq 9)$ perfect matching occurs. Figure \ref{fig:Route2}(c) shows the plot of the cross-correlation modularity measure $\langle C_M \rangle_r$ of Eq.~(\ref{eq:CMM}) versus $M_{syn}^{(inter)}$. Thus, for $M_{syn}^{(inter)} < {M_{syn}^{(inter)}}^*$ modular sparse synchronization with $\langle C_M \rangle_r<1$ emerges, while global sparse synchronization with $\langle C_M \rangle_r=1$ appears for $M_{syn}^{(inter)} > {M_{syn}^{(inter)}}^*$. In this way, with increasing $M_{syn}^{(inter)}$ the pacing degree between spikes increases monotonically thanks to the increase in the degree of effectiveness of global communication between spikes. Hence, $M_{syn}^{(inter)}$ plays only a constructive role to favor the pacing between spikes in sub-networks as well as the matching between the intra-modular dynamics of the sub-networks, in contrast to dual roles of $J_{inter}$ for the case of route I. Hence, unsynchronization does not appear. This constructive role of $M_{syn}^{(inter)}$ may be seen explicitly in Figs.~\ref{fig:Route2}(d1)-\ref{fig:Route2}(f2). We first consider the occupation degree which characterizes the sparseness degree of spike synchronization. For the case of modular synchronization, the sub-population occupation degree $\langle O_s^{(I)} \rangle_r$ in the sub-networks decreases a little with increasing $M_{syn}^{(inter)}$, while  $\langle O_s^{(I)} \rangle_r$ remains nearly constant for the case of global synchronization. On the other hand, as $M_{syn}^{(inter)}$ is increased the whole-population occupation degree $\langle O_w \rangle_r$ increases and approaches $\langle O_s^{(I)} \rangle_r$ because of decrease in the mismatching degree between the intra-modular dynamics of the sub-networks, and eventually when passing the threshold ${M_{syn}^{(inter)}}^*$ they become the same and then remain nearly constant with increasing $M_{syn}^{(inter)}$. Hence, the constant behavior of $\langle O_s^{(I)} \rangle_r$ and $\langle O_w \rangle_r$ for the case of global synchronization (which may occur because the average inhibition given to each neuron is the same for constant inter-modular coupling strength, independently of $M_{syn}^{(inter)}$) is in contrast to the monotonically-decreasing behavior of $\langle O_s^{(I)} \rangle_r$ and $\langle O_w \rangle_r$ for the case of route I [refer to Figs.~\ref{fig:SM}(a1)-\ref{fig:SM}(a4)]. Since $\langle O_s^{(I)} \rangle_r <1$ [i.e., only a small fraction of the total $L$ $(=10^3)$ neurons in the sub-population fire in each stripe], modular and global synchronization is sparse one. Next, we consider the pacing degree between spikes in the stripes. For both cases of modular and global sparse synchronization, with increasing $M_{syn}^{(inter)}$ both the sub- and the whole-population pacing degrees $\langle P_s^{(I)} \rangle_r$ and  $\langle P_w \rangle_r$ increase monotonically due to a constructive role of $M_{syn}^{(inter)}$ favoring the pacing between the spikes, in contrast to the bell-shaped behavior for the case of the route I [refer to Figs.~\ref{fig:SM}(b1)-\ref{fig:SM}(b4)]. Consequently, both the sub- and the whole-population statistical-mechanical spiking measures $M_s^{(I)}$ and $M_s^{(w)}$ (which are given by the products of the sub- and the whole-population occupation and pacing degrees) increase monotonically in both cases of modular and global sparse synchronization, which is also in contrast to the case of the route I [refer to Figs.~\ref{fig:SM}(c1)-\ref{fig:SM}(c4)]. To further understand the pacing degree between spikes in the stripes, we consider the sub-population spatial cross-correlation degree $\langle \langle C_l^{(I)} \rangle_l \rangle_r$ between neuronal pairs [given by double averaging of the spatial cross-correlation function $C_l^{(I)}$ of Eq.~(\ref{eq:SCC}) over all lengths $l$ and realizations]. Figure \ref{fig:Route2}(g) shows plots of $\langle \langle C_l^{(I)} \rangle_l \rangle_r$ (obtained via average over 20 realizations) versus $M_{syn}^{(inter)}$ for $I=$1, 2, and 3. Similar to the case of the sub-population pacing degree $\langle P_s^{(I)} \rangle_r$, $\langle \langle C_l^{(I)} \rangle_l \rangle_r$ also displays monotonic increasing behavior. Hence, the statistical-mechanical pacing degree between spikes seems to be associated with the microscopic spatial cross-correlation degree between neuronal pairs, like the case of route I.

\begin{figure}
\includegraphics[width=0.9\columnwidth]{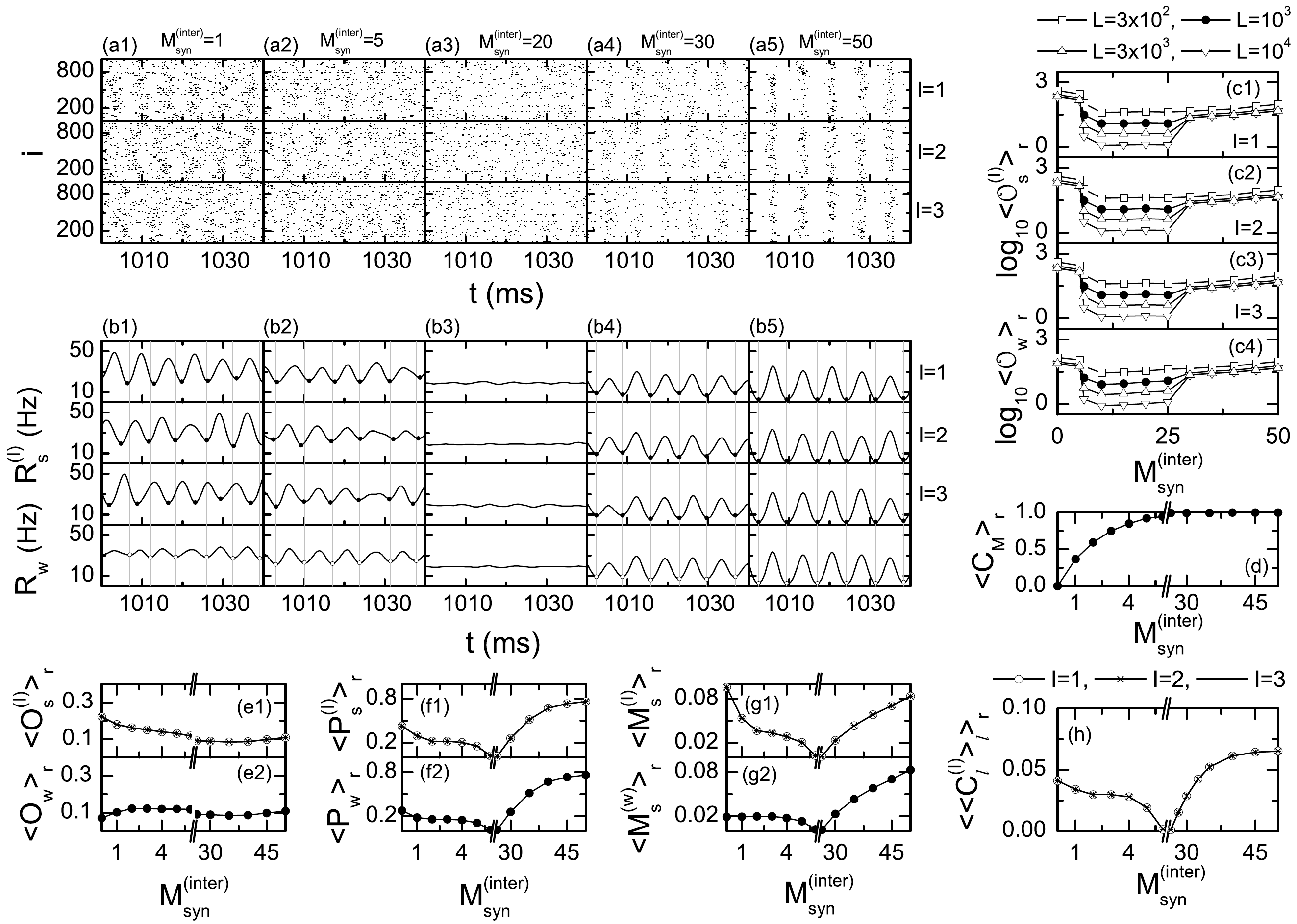}
\caption{Modular and global sparse synchronization along the route III with $J_{inter}=2500$ for the 1st case of intra-modular dynamics with $p_{rewiring}=0.25$.
Raster plots of neural spikes in the sub-networks ($I=1,2,3)$ for $M_{syn}^{(inter)}=$ (a1) 1, (a2) 5, (a3) 20, (a4) 30 and (a5) 50. Instantaneous sub- and whole-population spike rate kernel estimates $R_s^{(I)}(t)$ $(I=1,2,3)$ and $R_w(t)$ for $M_{syn}^{(inter)}=$ (a1) 1, (a2) 5, (a3) 20, (a4) 30 and (a5) 50. Vertical gray lines pass minima of $R_w(t)$, and the minima of $R_s^{(I)}(t)$ [$R_w(t)$] are represented by solid (open) circles. (c1)-(c4) Plots of the sub- and the whole-population order parameters $\log_{10}{\langle {\cal{O}}_s^{(I)} \rangle_r}$ ($I=1,2,3$) and $\log_{10}{\langle {\cal{O}}_w \rangle_r}$ versus $M_{syn}^{(inter)}$. In (d)-(h), break symbols are given in the unsynchronization regions; the left (right) parts of the break symbols correspond to the regions of modular (global) synchronization. (d) Plot of the cross-correlation modularity measure $\langle C_M \rangle_r$ versus $M_{syn}^{(inter)}$. (e1)-(e2) Plots of the sub- and the whole-population occupation degrees ${\langle O_s^{(I)} \rangle}_r$ and  ${\langle O_w \rangle}_r$ versus $M_{syn}^{(inter)}$. (f1)-(f2) Plots of the sub- and the whole-population pacing degrees ${\langle P_s^{(I)} \rangle}_r$ and  ${\langle P_w \rangle}_r$ versus $M_{syn}^{(inter)}$. (g1)-(g2) Plots of the sub- and the whole-population statistical-mechanical spiking measures ${\langle M_s^{(I)} \rangle}_r$ and ${\langle M_s^{(w)} \rangle}_r$ versus $M_{syn}^{(inter)}$. (h) Plot of the spatial cross-correlation degree ${\langle {\langle C_l^{(I)} \rangle}_l \rangle}_r$ versus $M_{syn}^{(inter)}$.
}
\label{fig:Route3}
\end{figure}

We also investigate emergence of modular and global sparse synchronization by increasing $M_{syn}^{(inter)}$ along the route III for $J_{inter}=2500$ (which is much larger than that for the case of route II). Unlike the case of the route II, for small $M_{syn}^{(inter)}$ a destructive effect to decrease the pacing degree between spikes occurs due to strong inhibition for $J_{inter}=2500$, and hence when passing a lower threshold ${M_{syn,l}^{(inter)}}^*$ $(\simeq 6)$ a transition from modular sparse synchronization to unsynchronization occurs. However, with further increase in $M_{syn}^{(inter)}$ a constructive effect of ${M_{syn}^{(inter)}}$ to favor the pacing between spikes and the matching between the intra-modular dynamics of sub-networks overcomes the destructive effect of strong inhibition. Consequently, a transition to global sparse synchronization occurs when passing a higher threshold ${M_{syn,h}^{(inter)}}^*$ $(\simeq 26)$. These results are well shown in Figs.~\ref{fig:Route3}(a1)-\ref{fig:Route3}(h). The raster plots of spikes in the three sub-populations for $M_{syn}^{(inter)}=$1, 5, 20, 30, and 50 are shown in Figs.~\ref{fig:Route3}(a1)-\ref{fig:Route3}(a5), respectively. The corresponding ISPSR and IWPSR kernel estimates, $R_s^{(I)}(t)$ and $R_w(t)$, for $M_{syn}^{(inter)}=$1, 5, 20, 30, and 50 are also shown in Figs.~\ref{fig:Route3}(b1)-\ref{fig:Route3}(b5), respectively. For each $I$th sub-population, sparse stripes are formed in the raster plot and $R_s^{(I)}(t)$ shows a regular oscillation, except for the unsynchronized case of $M_{syn}^{(inter)}=20$ where spikes are scattered without forming stripes in the raster plot and $R_s^{(I)}(t)$ is nearly stationary. Figures \ref{fig:Route3}(c1)-\ref{fig:Route3}(c4) show the sub- and the whole-population order parameters ${\cal{O}}_s^{(I)}$ and ${\cal{O}}_w$ which determine a threshold for the synchronization-unsynchronization transition. In the region of ${M_{syn,l}^{(inter)}}^* <M_{syn}^{(inter)} < {M_{syn,h}^{(inter)}}^*$, both ${\cal{O}}_s^{(I)}$ and ${\cal{O}}_w$ tend to zero in the thermodynamic limit of $L \rightarrow \infty$, and hence unsynchronized states appear due to a destructive effect of strong inhibition. On the other hand, for $M_{syn}^{(inter)} < {M_{syn,l}^{(inter)}}^*$ or $M_{syn}^{(inter)} > {M_{syn,h}^{(inter)}}^*$, the values of $\langle {\cal{O}}_s^{(I)} \rangle_r$ and $\langle {\cal{O}}_w \rangle_r$ become saturated to non-zero limit values for large $L$, and hence synchronized states exist. Particularly, for $M_{syn}^{(inter)} > {M_{syn,h}^{(inter)}}^*$ sparsely synchronized states appear due to a constructive effect of ${M_{syn}^{(inter)}}$ favoring the pacing between spikes. The type of sparse synchronization may be determined in terms of the cross-correlation modularity measure $C_M$ of Eq.~(\ref{eq:CMM}) which is shown in Fig.~\ref{fig:Route3}(d). For $M_{syn}^{(inter)} < {M_{syn,l}^{(inter)}}^*$ modular sparse synchronization with $\langle C_M  \rangle_r <1$ (i.e., some mismatching between the intra-modular dynamics of sub-networks) emerges, while global sparse synchronization with $\langle C_M \rangle_r =1 $ (i.e., perfect matching between the intra-modular dynamics of sub-networks) appears for $M_{syn}^{(inter)} > {M_{syn,h}^{(inter)}}^*$. The degree of synchronization is also measured in terms of the occupation degrees, the pacing degrees, and the statistical-mechanical spiking measures in the sub- and the whole-populations, which are shown in Figs.~\ref{fig:Route3}(e1)-\ref{fig:Route3}(g2). We first consider the case of modular sparse synchronization. As $M_{syn}$ is increased, both the occupation degree $\langle O_s^{(I)} \rangle_r$ and the pacing degree $\langle P_s^{(I)} \rangle_r$ in the sub-networks decrease due to a destructive effect of strong inhibition for $J_{inter}=2500$. In the whole-population, with increasing $M_{syn}$ the occupation degree $\langle O_w \rangle_r$ increases and approaches $\langle O_s^{(I)} \rangle_r$ because of decrease in the mismatching degree between the intra-modular dynamics of sub-networks, and the pacing degree $\langle P_w \rangle_r$ decreases like the case of $\langle P_s^{(I)} \rangle_r$. Thus, both the sub- and the whole-population statistical-mechanical spiking measures $M_s^{(I)}$ and $M_s^{(w)}$ (which are given by the products of the sub- and the whole-population occupation and pacing degrees) decrease as $M_{syn}^{(inter)}$ increases. On the other hand, for the case of global sparse synchronization which is similar to the case of route II, the constructive effect of $M_{syn}^{(inter)}$ favoring the pacing between spikes dominates. Consequently, with increasing $M_{syn}^{(inter)}$ both $\langle P_s^{(I)} \rangle_r$ and $\langle P_w \rangle_r$ increase monotonically, while both $\langle O_s^{(I)} \rangle_r$ and $\langle O_w \rangle_r$ remains nearly constant because the average inhibition given to each neuron is the same for constant inter-modular coupling strength $J_{inter}$, independently of $M_{syn}^{(inter)}$. Consequently, as $M_{syn}^{(inter)}$ is increased, the sub- and the whole-population statistical-mechanical spiking measures $M_s^{(I)}$ and $M_s^{(w)}$ show a monotonic increase. As in the case of route II, the modular and global synchronization is sparse one because $\langle O_s^{(I)} \rangle_r$ is much less than unity. Furthermore, the statistical-mechanical pacing degree between spikes in the sub-population seems to be associated with the spatial cross-correlation degree $\langle \langle C_l^{(I)} \rangle_l \rangle_r$ between neuronal pairs (obtained through average 20 realizations), which is shown in Fig.~\ref{fig:Route3}(h). For the case of modular sparse synchronization (i.e., $M_{syn}^{(inter)} < {M_{syn,l}^{(inter)}}^*$), with increasing $M_{syn}^{(inter)}$ $\langle \langle C_l^{(I)} \rangle_l \rangle_r$ decreases monotonically due to a destructive role of strong inhibition, while for the case of global sparse synchronization (i.e., $M_{syn}^{(inter)} > {M_{syn,h}^{(inter)}}^*$), $\langle \langle C_l^{(I)} \rangle_l \rangle_r$ exhibits a monotonic increase because of a constructive role of $M_{syn}^{(inter)}$.

\subsection{2nd and 3rd Cases of Intra-Modular Dynamics}
\label{subsec:Case23}
To further examine dependence of the inter-modular connection effect on the type of intra-modular dynamics, we consider the 2nd and the 3rd cases of the intra-modular dynamics: (2) unsynchronized in the absence of inter-modular coupling in all identical sub-networks with $p_{rewiring}=0.05$ and (3) non-identical sub-networks where in the absence of inter-modular coupling, the 1st sub-network with $p_{rewiring}=0.25$ is synchronized, the 2nd sub-network with $p_{rewiring}=0.15$ is also synchronized, but the 3rd sub-network with $p_{rewiring}=0.05$ is unsynchronized. Figure \ref{fig:SD23}(a) shows the state diagram in the $J_{inter}$-$M_{syn}^{(inter)}$ plane for the 2nd case of intra-modular dynamics. This state diagram is similar to that for the 1st case in Fig.~\ref{fig:SD1}(g), except for the appearance of ``L''-shaped region of unsynchronization for small $J_{inter}$ or $M_{syn}^{(inter)}$. Beyond the unsynchronized region, modular and global sparse synchronization appears in the gray and the dark gray regions, respectively. For large $J_{inter}$ $(> 1402)$ unsynchronization also occurs for large $J_{inter}$ $(> 1402)$ between the modular and the global synchronization. We also make an intensive investigation of emergence of modular and global sparse synchronization by changing $J_{inter}$ along the route of $M_{syn}^{(inter)}=20$ (corresponding to the 1st route I for the 1st case). As in the 1st case, we obtain the raster plots of spikes in the three sub-populations and the corresponding ISPSR and IWPSR kernel estimates, $R_s^{(I)}(t)$ and $R_w(t)$ for representative values of $J_{inter}=$50, 200, 600, 1000, and 2000; for brevity associated figures are not presented. Unlike the 1st case 1 of the intra-modular dynamics, for small $J_{inter}$ (=50) unsynchronization occurs because of small $p_{rewiring}$ (=0.05). For this case, the clustering coefficient is high, and hence partial stripes (indicating local clustering of spikes) seem to appear in the raster plots of spikes. Thus, the raster plots show zigzag patterns intermingled with partial stripes with diverse inclinations and widths, and hence spikes become difficult to keep pace with each other. Consequently, $R_s^{(I)}(t)$ and $R_w(t)$ become nearly stationary. However, as $J_{inter}$ is increased and passes a lower threshold $J_{inter,l}^*$ $(\simeq 187)$, the inter-modular coupling strength $J_{inter}$ plays a constructive role to favor the pacing between spikes in each sub-network, and synchronized states with regularly-oscillating $R_s^{(I)}(t)$ and $R_w(t)$ appear for $J_{inter}=$ 200, 600, and 1000. On the other hand, for large $J_{inter}$, due to strong inhibition $J_{inter}$ plays a destructive role to spoil the pacing between sparse spikes. Hence, when passing a higher threshold $J_{inter,h}^*$ $(\simeq 1402)$, a transition to unsynchronization occurs, as shown for $J_{inter}=2000$. For this case, sparse spikes in the raster plots in each sub-network are scattered without forming stripes, and hence both $R_s^{(I)}(t)$ and $R_w(t)$ become nearly stationary (i.e., every sub-network exhibits an unsynchronized state). Similar to the 1st case, one can consider additional routes with fixed values of $J_{inter}$ (e.g., 750 and 2300). As shown in the state diagram of Fig.~\ref{fig:SD23}(a), unsynchronization occurs for small $M_{syn}^{(inter)}$, in contrast to the 1st case. However, when passing a threshold $M_{syn}^{(inter)*}$ modular synchronization appears, and then the population behaviors are similar to those for the 1st case.

Finally, we consider the 3rd case of non-identical sub-networks where in the absence of inter-modular coupling, the 1st sub-network with $p_{rewiring}=0.25$ is synchronized, the 2nd sub-network with $p_{rewiring}=0.15$ is also synchronized, but the 3rd sub-network with $p_{rewiring}=0.05$ is unsynchronized. Thanks to a constructive role of $J_{inter}$ favoring the pacing between spikes, a transition to synchronization occurs in the 3rd sub-network when passing a lower threshold $J_{inter,l}^*$. With increasing $M_{syn}^{(inter)}$, the value of $J_{inter,l}^*$ decreases due to a constructive effect of $M_{syn}^{(inter)}$ to favor the pacing between spikes. For $J < J_{inter,l}^*$, the 3rd sub-network is still unsynchronized, while the 1st and the 2nd sub-systems are synchronized. As $J_{inter}$ passes $J_{inter,l}^*$, the 3rd sub-network becomes synchronized, and then modular synchronization occurs due to mismatching between the intra-modular dynamics of sub-networks. Here, we consider the case of $J_{inter} > J_{inter,l}^*$. Figure \ref{fig:SD23}(b) shows the state diagram in the $J_{inter}$-$M_{syn}^{(inter)}$ plane for the case of $J_{inter} \geq 1$ and $M_{syn}^{(inter)} \geq 1$ (i.e., the region where the 1st and the 2nd sub-networks are synchronized but the 3rd sub-network is unsynchronized is not shown). We note that this state diagram is nearly the same as that for the case 1 in Fig.~\ref{fig:SD1}(g). Modular sparse synchronization occurs in the ``L''-shaped gray region, while global synchronization appears in the dark gray region. Unsynchronization also occurs for large $J_{inter}$ $(> 1371)$ between modular and global synchronization. When compared with the 1st case of the intra-modular dynamics, the regions of modular synchronization and unsynchronization are a little enlarged, while the region of global synchronization is somewhat contracted. We make an intensive study on appearance of modular and global synchronization by increasing $J_{inter}$ from the threshold $J_{inter,l}^* (\simeq 0.2)$ for the 3rd sub-network along the route of $M_{syn}^{(inter)}=20$ (corresponding to the 1st route I for the 1st case). Similar to the 1st case, we obtain the raster plots of spikes in the three sub-populations and the corresponding ISPSR and IWPSR kernel estimates, $R_s^{(I)}(t)$ and $R_w(t)$ for representative values of $J_{inter}=$10, 100, 500, 1000, and 2000; for brevity associated figures are not presented. For $J_{inter}=10$, due to a constructive role of $J_{inter}$ favoring the pacing between spikes, sparse stripes are formed in each $I$th sub-population $(I=1, 2, 3)$. However, in contrast to the case 1 of identical intra-modular dynamics, the smearing degree of stripes is different, depending on the sub-population. The stripes for the 1st sub-population (with $p_{rewiring}=0.25$) are relatively clear, while the stripes in the other 2nd and 3rd sub-populations (with $p_{rewiring}=0.15$ and 0.05, respectively) are more and more smeared. Hence, the amplitudes of the regularly-oscillating $R_s^{(I)}(t)$ decrease as $I$ is increased. As $J_{inter}$ is further increased (e.g., $J_{inter}$=100 and 500), the pacing degree of spikes in the stripes increases for each sub-population, although the stripes become more sparse. However, for large $J_{inter}$ (e.g., $J_{inter}=1000$), due to a destructive role of $J_{inter}$ the pacing degree of spikes in the stripes begins to decrease. Eventually, when passing a higher threshold $J_{inter,h}^*$ $(\simeq 1372)$ a transition to unsynchronization occurs. Thus, for $J_{inter}=2000$, spikes are scattered in the raster plot, and both $R_s^{(I)}(t)$ and $R_w(t)$ are nearly stationary. To study the effect of $M_{syn}^{(inter)}$, one may also consider another routes with fixed values of $J_{inter}$ (e.g., 750 and 2300). With increasing $M_{syn}^{(inter)}$ along these routes, population behaviors are similar to those for the 1st case, as shown in the state diagram of Fig.~\ref{fig:SD23}(b).

\section{Summary}
\label{sec:SUM}
\begin{figure}
\includegraphics[width=0.8\columnwidth]{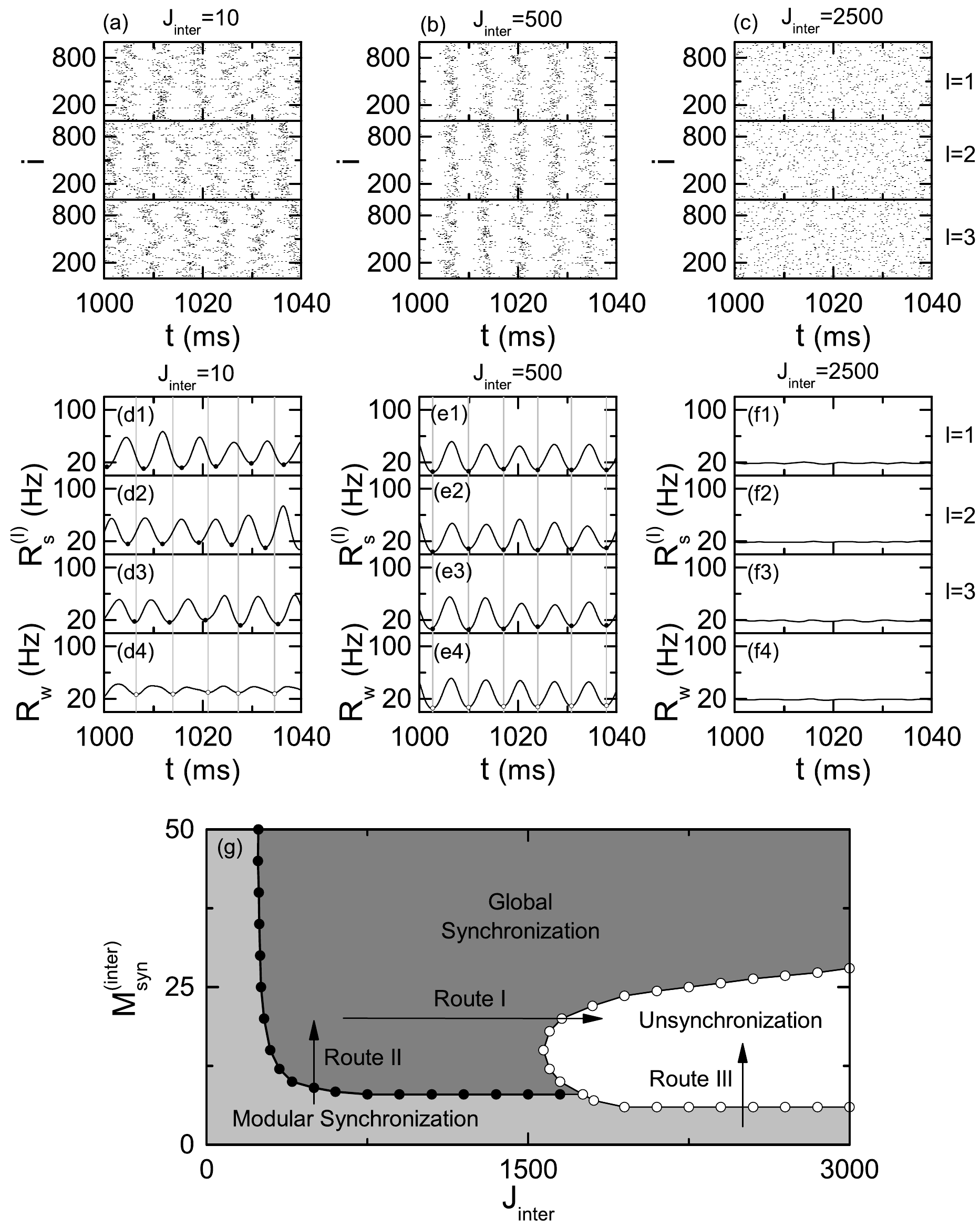}
\caption{Modular and global sparse synchronization for the 1st case of intra-modular dynamics with $p_{rewiring}=0.25$. $M_{syn}^{(inter)}=20$ in (a)-(f4).
Raster plots of neural spikes in the sub-networks ($I=1,2,3)$ for $J_{inter}=$ (a)10, (b) 500, and (c) 2500. Instantaneous sub- and whole-population spike rate kernel estimates, $R_s^{(I)}(t)$ $(I=1,2,3)$ and $R_w(t)$:
$J_{inter}=$ (d1)-(d4) 10, (e1)-(e4) 500, and (f1)-(f4) 2500. Vertical gray lines for $J_{inter}=10$ and $500$ pass minima of $R_w(t)$, and the minima of $R_s^{(I)}(t)$ [$R_w(t)$] are denoted by solid (open) circles. (g) State diagram in the $J_{inter}$-$M_{syn}^{(inter)}$ plane.
}
\label{fig:SD1}
\end{figure}

\begin{figure}
\includegraphics[width=0.9\columnwidth]{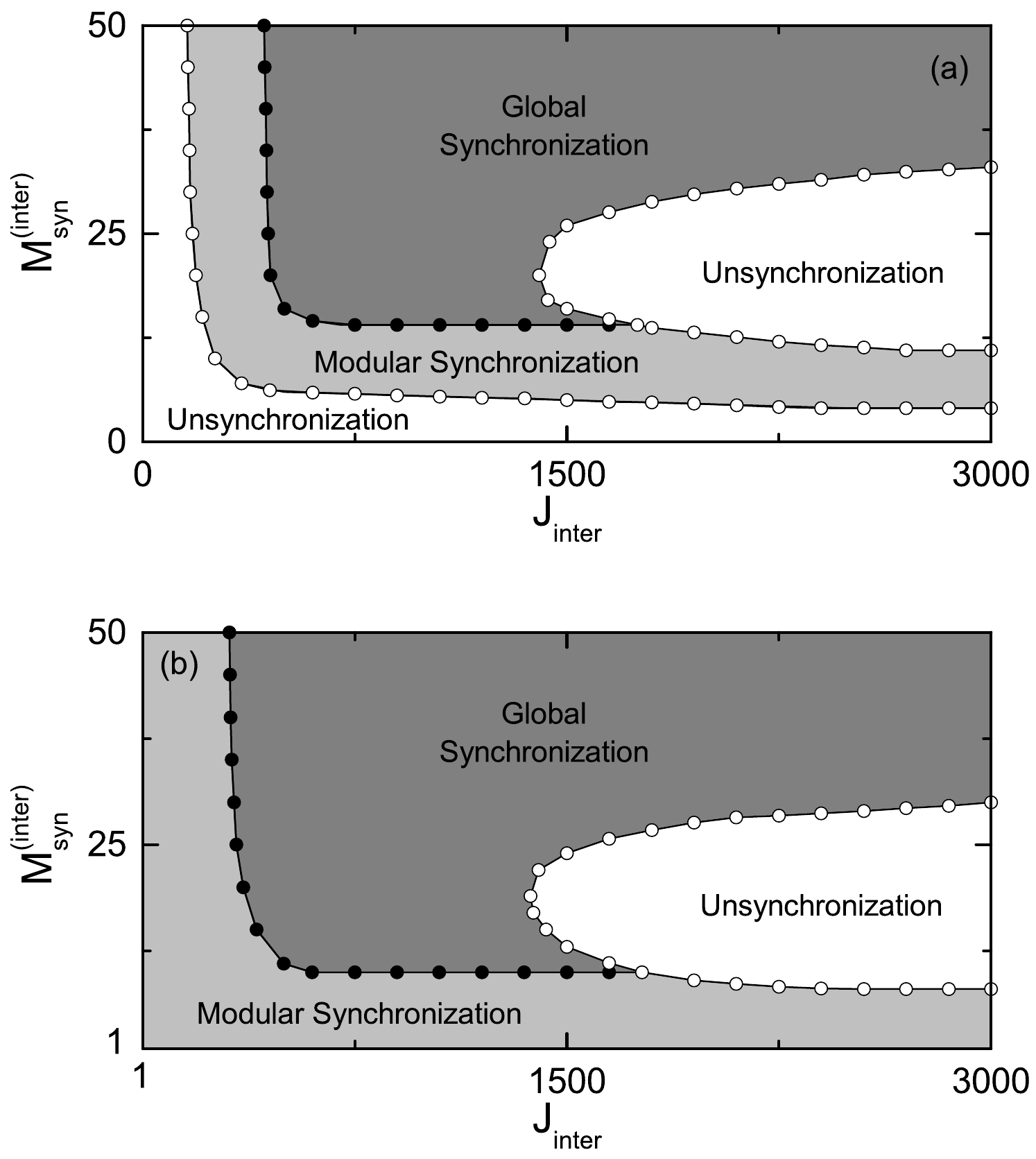}
\caption{(a) State diagram in the $J_{inter}$-$M_{syn}^{(inter)}$ plane for the 2nd case of the intra-modular dynamics (i.e., in the absence of inter-modular coupling, identically unsynchronized sub-networks with $p_{rewiring}=0.05$). (b) State diagram in the $J_{inter}$-$M_{syn}^{(inter)}$ plane for the 3rd case of the intra-modular dynamics (i.e., in the absence of inter-modular coupling, synchronized in the 1st sub-network with
$p_{rewiring}$ = 0.25, less synchronized in the 2nd sub-network with $p_{rewiring}$ = 0.15, and unsynchronized in the 3rd sub-network $p_{rewiring}$ = 0.05).
}
\label{fig:SD23}
\end{figure}

Sparsely inter-connected modular structures are found in both mammalian brain anatomical networks and human brain functional networks, as in other complex systems such as social, technological, and biological networks. Modular organization of the brain network shows the anatomical substrate for segregation of the brain into specialized sub-regions with particular functional tasks. These specialized informations of different features are also integrated to produce new useful information as a whole. In this way, the brain network is organized via the interplay between segregation (specialization) and integration (binding). We note that these real brains, composed of sparsely linked clusters, are far more complex than minimal non-clustered models such as small-world and scale-free networks. To take into consideration the modular structure of the real brain, we considered clustered small-world networks of inhibitory FS interneurons, and investigated the effect of inter-modular connection on emergence of sparsely synchronized rhythms at the sub- and whole-population levels by employing diverse realistic measures. By changing both the inter-modular coupling strength $J_{inter}$ and the average number of inter-modular links per interneuron $M_{syn}^{(inter)}$, we made intensive study on emergence of sparsely synchronized population states along the three routes in the $J_{inter}-$$M_{syn}^{(inter)}$ plane for the 1st case of the intra-modular dynamics: (1) synchronized in all identical sub-networks. Consequently, two kinds of sparse synchronization such as modular and global synchronization have been found to appear, in contrast to the case of non-clustered networks. Our main findings are that the type and degree of sparse synchronization depend on the inter-modular parameters, $J_{inter}$ and $M_{syn}^{(inter)}$. For lower values $J_{inter}$ acts to favor the pacing between spikes, while for higher values it tends to spoil the pacing between spikes due to strong inhibition. On the other hand, with increasing $M_{syn}^{(inter)}$ it acts to monotonically increase the pacing between spikes, which results from increase in the degree of effectiveness of global communication between spikes. To examine dependence on the intra-dynamics in the sub-networks, we have also considered two other cases for the intra-dynamics: (2) unsynchronized in all identical sub-networks and (3) synchronized/unsynchronized in non-identical sub-networks. Figures \ref{fig:SD1}(g), \ref{fig:SD23}(a) and \ref{fig:SD23}(b) show the state diagrams, representing main features on the population states, for the 1st, 2nd, and 3rd cases, respectively. For the 2nd case, an ``L''-shaped region of unsynchronization appears for small $J_{inter}$ or $M_{syn}^{(inter)}$ because the intra-modular dynamics in all sub-networks are unsynchronized in the absence of inter-modular coupling. Beyond this ``L''-shaped region, the structure of the state diagram is similar to that for the 1st case. In the 3rd case, its state diagram is nearly the same as that for the 1st case. Due to non-identicalness of sub-networks, modular synchronization persists a little more, and hence the region of modular synchronization is a little enlarged when compared to the 1st case. Moreover, in the case of modular synchronization the pacing degree of spikes varies depending on the sub-networks, in contrast to the 1st case of identical sub-networks. From the results in these three cases, it follows that the effect of inter-modular connections seems to be essentially the same, independently of the type of intra-modular dynamics in sub-networks. Finally, since changes in the coupling strengths and the synaptic connections are closely interwoven with the brain plasticity \cite{plastic}, we expect that our results on the inter-modular connection effect in modular networks might have implications for the role of the brain plasticity in some functional behaviors related to population synchronization. However, explicit study on the inter-relation between inter-modular connection, population synchronization, and brain plasticity is beyond our present subject, and it is left as a future work.

\begin{acknowledgments}
This research was supported by Basic Science Research Program through the National Research Foundation of Korea(NRF) funded by the Ministry of Education (Grant No. 2013057789).
\end{acknowledgments}

\newpage
\begin{table}
\caption{Parameter values used in our computations for Figs.~\ref{fig:SD}-\ref{fig:SD23}; units of the capacitance, the potential, the current, and the time are pF, mV, pA, and ms, respectively.}
\label{tab:Parm}
\begin{ruledtabular}
\begin{tabular}{llllll}
(1) & \multicolumn{5}{l}{Single Izhikevich FS Interneurons \cite{Izhi3}} \\
& $C=20$ & $v_r=-55$ & $v_t=-40$ & $v_p=25$ & $v_b=-55$  \\
& $k=1$ & $a=0.2$ & $b=0.025$ & $c=-45$ & $d=0$  \\
\hline
(2) & \multicolumn{5}{l}{External Stimulus to Izhikevich FS Interneurons} \\
& $I_{DC} = 1500$ & $D=500$ \\
\hline
(3) & \multicolumn{5}{l}{Inhibitory GABAergic Synapse \cite{Sparse3}} \\
& $\tau_l=1$ & $\tau_r=0.5$ & $\tau_d=5$ & $V_{syn}=-80$ \\
\hline
(4) & \multicolumn{5}{l}{Intra-modular Coupling in Small-world Sub-networks} \\
& $M_{syn}^{(intra)}=50$ & $J_{intra}=1400$ & $p_{rewiring}:$ Varying \\
\hline
(5) & \multicolumn{5}{l}{Inter-modular Connection} \\
& $M_{syn}^{(inter)}:$ Varying & $J_{inter}:$ Varying
\end{tabular}
\end{ruledtabular}
\end{table}

\newpage

\newpage

\newpage

\newpage

\newpage

\newpage

\newpage

\newpage

\newpage

\newpage

\end{document}